\documentclass[a4paper,pre,reqno,superscriptaddress,showpacs,twocolumn]{revtex4-1}
\usepackage{graphicx,color}
\usepackage{dcolumn}
\usepackage{epsfig,changes}
\usepackage{float}
\usepackage[centertags]{amsmath}
\usepackage{amsfonts,amsmath}
\usepackage{euscript}
\usepackage{amssymb}
\usepackage{amsthm}
\usepackage{mathptmx}
\usepackage{newlfont}
\usepackage{mathrsfs}
\usepackage{subfigure}

\newcommand{\opunit}{\text{1}\kern-0.22em\text{l}}

\newcommand{\eg}{\textit{e.g.}}
\newcommand{\ie}{\textit{i.e.}}

\newcommand{\id}{\textrm{d}}

\def\bea{\begin{eqnarray}}
\def\eea{\end{eqnarray}}
\def\ba{\begin{array}}
\def\ea{\end{array}}
\def\n{\nonumber}

\def\la{\langle}
\def\ra{\rangle}

\begin{document}
 
\title{Run-and-tumble particles in two-dimensions : Marginal position distributions}
\author{Ion Santra}
\author{Urna Basu}
\author{Sanjib Sabhapandit}
\affiliation{Raman Research Institute, Bengaluru 560080, India}

\begin{abstract}
We study a set of Run-and-tumble particle (RTP)  dynamics in two spatial dimensions. In the first case of the orientation $\theta$ of the particle can assume a set of $n$ possible discrete values while in the second case $\theta$ is a continuous variable. We calculate exactly the marginal position distributions for $n=3,4$ and the continuous case and show that in all the cases the RTP shows a cross-over from a ballistic to diffusive regime. The ballistic regime
is a typical signature of the active nature of the systems and is characterized by non-trivial position distributions which depends on the specific model. We also show that, the signature of activity at long-times  can be found in the atypical fluctuations which we also characterize by computing the large deviation functions explicitly.
\end{abstract}

\maketitle

\section{Introduction}

Active particles are self-propelled systems which can generate dissipative, persistent motion by extracting energy from their surroundings at the individual particle level \cite{Romanczuk,soft,BechingerRev,Ramaswamy2017,Marchetti2017,Schweitzer}. Numerous examples of active systems can be found in nature, ranging from bacterial motion \cite{Berg2004,Cates2012}, cellular and tissue motility \cite{tissue1,tissue2}  to granular matter \cite{gran1,gran2},  fish-schools \cite{Vicsek,fish} and flock of birds \cite{flocking1, flocking2}. The inherent nonequilibrium nature of active particles lead to many remarkable features which are strikingly different than their equilibrium counterparts.  For example, interacting active particles show a plethora of novel emergent collective behaviour like mobility induced phase separation \cite{separation1, separation2, separation3}, clustering \cite{cluster1,cluster2,evans} and absence of well defined pressure \cite{Kardar2015}. On the other hand, single active particles also show a wide range of intriguing features like non-Boltzmann stationary state, clustering near the boundaries of the confining region \cite{Solon2015, Potosky2012, ABP2019, RTP_trap, Malakar2019, Takatori} and unusual relaxation and persistence properties \cite{RTP_free, ABP2018, Singh2019, Franosch2016, Franosch2018}.

An important focus of the theoretical attempt to understand and characterize the behaviour of active particles is the study of minimal statistical models of such systems.  Run-and-Tumble particles  (RTP)  is one such model which describes the motion of an overdamped particle which moves or `runs' with a constant speed along an internal spin direction. This  internal direction also changes stochastically which results in the `tumble' of the active particle. Originally introduced as a model for bacterial motion, RTP dynamics has emerged as one of the fundamental non-equilibrium toy models for studying many aspects of active particle dynamics. The simplest and most studied version is the one dimensional RTP where the internal spin can assume two possible directions \cite{RTP_free, RTP_trap}. RTP with multiple internal states have also been studied \cite{Maes2018,Seifert2016}. Such multi-state models might arise naturally from multi-particle scenarios \cite{Majumdar2019, gel} or higher spatial dimension \cite{3st-RTP2019}.

 Behaviour of RTP in higher spatial dimension is an interesting topic in itself and have been studied much in the past few years\cite{Solon2015,Swimmer_2d,RTP_swimmer2d,RTP_2_3d,Stadje,Martens2012,Active2d}. RTP with rotational diffusion and arbitrary run-time distributions have also been studied in the context of maximal diffusivity \cite{RTParbitruntime1,RTParbitruntime2} and minimal navigation strategies in presence of an external field \cite{Markovrobots}. However, not much analytical results are available regarding position distribution of RTP in higher dimensions except Refs.~\cite{Stadje,Martens2012,Active2d}. In this article we study a set of RTP dynamics in two spatial dimensions (2D). 

 In 2D, the internal spin direction can be uniquely specified by an angle $\theta,$ which can take either discrete or continuous values.  
 We consider two different classes of RTPs  where  $\theta$ assumes  (i)  $n$ discrete directions in space and  (ii)  continuous values in the range $[0,2\pi].$ We compute the exact marginal position distributions for 
$n=3,4,$ and the continuous case. We find that strong signatures of activity are seen in the short-time regime in the form of spatial anisotropy and/or ballistic nature of the motion. We also show that, in the long-time regime, while the typical fluctuations in position are characterized by Gaussian distributions for all the models, the atypical fluctuations still contain signatures of activity, which we characterize with the help of large deviation functions. 
 
 The paper is organized as follows. In the next section we introduce the models and present a brief summary of our main results. Sections \ref{sec:3st} and \ref{sec:4st} are devoted to the study of the cases $n=3$ and $n=4,$ respectively. We focus on the continuous-$\theta$ model in Sec.~\ref{sec:cont}. We conclude with some general remarks and a few open questions in Sec.~\ref{sec:conclusions}. 

\section{Models and Results}

Let us consider an overdamped particle moving on the two-dimensional $x-y$ plane. The particle moves with a constant speed $v_0$ along some internal direction or `spin' described by an angle $\theta.$  
The Langevin equations governing the time evolution of the position $ (x (t) ,y (t) ) $ are given by,
\bea
\dot x (t)  = v_0 \cos \theta (t)  \cr
\dot y (t)  = v_0 \sin \theta (t). 
\label{eq:dynamics}
\eea
The orientation $\theta$ itself changes stochastically which gives rise to the `active' nature of the motion. In this article, we consider two different kinds of dynamics for $\theta.$ \\

\noindent {I. \bf $n$-state model:}  In this case, $\theta$ can have $n$ possible discrete values $\theta=0,2\pi/n, 4\pi/n, \cdots  (n-1) 2\pi/n$ and evolves following a jump process -- the orientation of the particle changes by an amount $\pm 2\pi/n$  (\ie, the spin rotates either clockwise or anti-clockwise)  with rate $\gamma/2.$ 
 
The $\theta$-dynamics is independent of the position degree of freedom, and is nothing but a symmetric continuous time random walk on a one dimensional ring with $n$ sites with jump rate $\gamma/2.$  It is straightforward to calculate (see Appendix \ref{sec:app_theta} for the details) the corresponding propagator $P(\theta ,t| \theta_0,0)$, \ie, the probability that the orientation is $\theta$ at time $t,$ starting from $\theta_0$ at time $t=0,$ and it is given by,
\bea
P(\theta ,t | \theta_0,0) = \frac 1n \sum_{k=0}^{n-1} e^{i k (\theta - \theta_0)}e^{-\gamma t\big(1- \cos \frac{2 \pi k}n \big)}. \label{eq:th_prop}
\eea
Note that the $n \to \infty$ limit, with a rescaling $\gamma \propto n^2,$ yields the active Brownian motion. On the other hand, for any finite $n,$ at large-times $t \to \infty$ each of the $\theta$-values become equally likely.

In the following, we study the cases $n=3$ and $4$ in details and compute the marginal position distribution of the RTP analytically. We assume that, at time $t=0,$ the particle starts from the origin $x=y=0$ and the spin can be oriented along any of the possible $n$ directions with equal probability $\frac 1n$.\\
 
\noindent {II. \bf Continuous model:} The second case is where $\theta$ can take any real value in the range $[0,2\pi].$  At any time $t,$ with rate $\gamma,$ $\theta$ can change to a different value $\theta'$ distributed uniformly in $[0,2\pi].$ In this case also, we can immediately write down the propagator,
\bea
P(\theta,t | \theta_0,0) = e^{-\gamma t} \delta(\theta - \theta_0) +  (1-e^{-\gamma t}) \frac 1{2 \pi}.
\label{cont1}
\eea
Here the first term corresponds to the scenario where $\theta$ has not flipped up to time $t$ and the second term corresponds to at least one flip. 
Note that, this continuous model is not the $n \to \infty$ limit of the discrete model introduced before.
In Sec.~\ref{sec:cont} we compute the position distribution of this continuous-$\theta$ RTP.  Once again,  we consider the initial condition $x=y=0$ at time $t=0$ and the initial orientation $\theta_0$ is chosen from the uniform distribution 
\bea
\mu (\theta_0) = \frac{1}{2\pi},\forall \theta\in[0,2\pi].
\eea

\begin{figure*}[t]
 \includegraphics[width=5 cm]{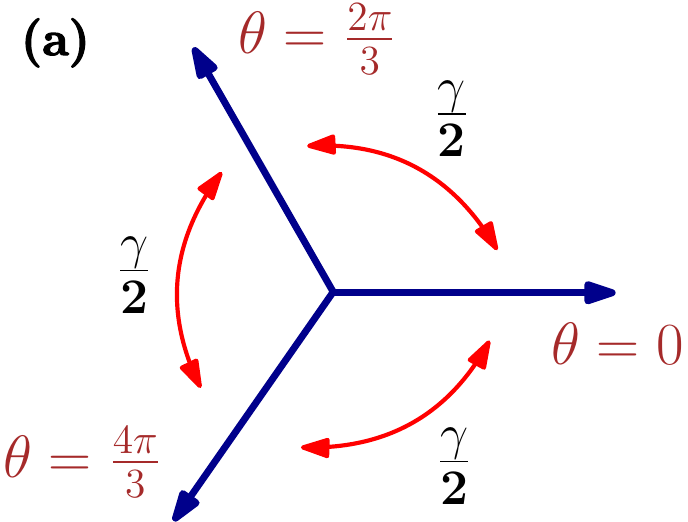}
 \includegraphics[width=5 cm]{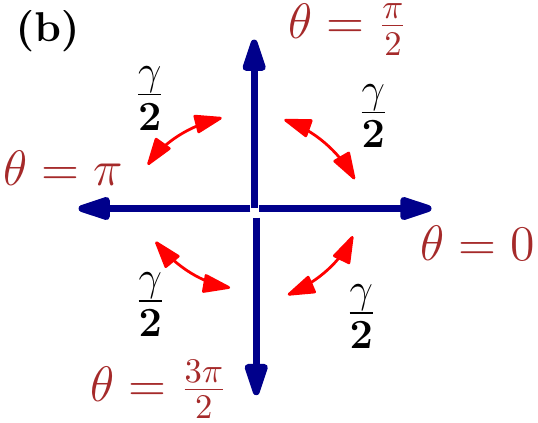}  
 \includegraphics[width=4.8 cm]{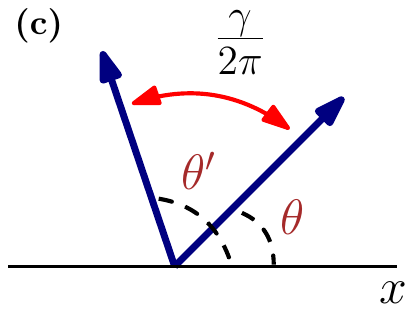}
 \vspace*{0.5 cm}
 \includegraphics[width=5 cm]{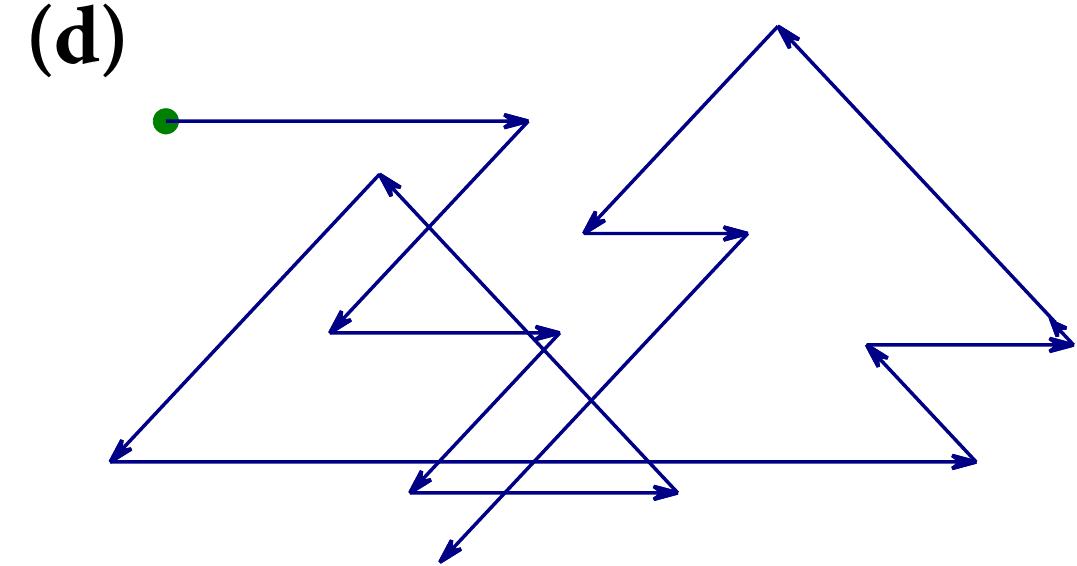}
 \includegraphics[width=5 cm]{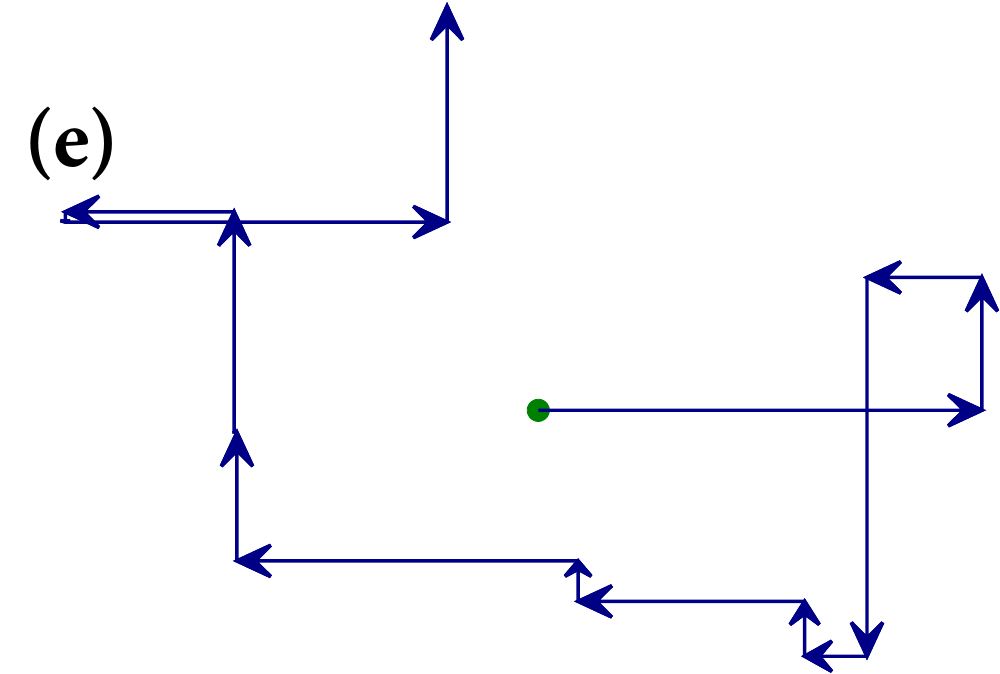}
  \includegraphics[width=4.8 cm]{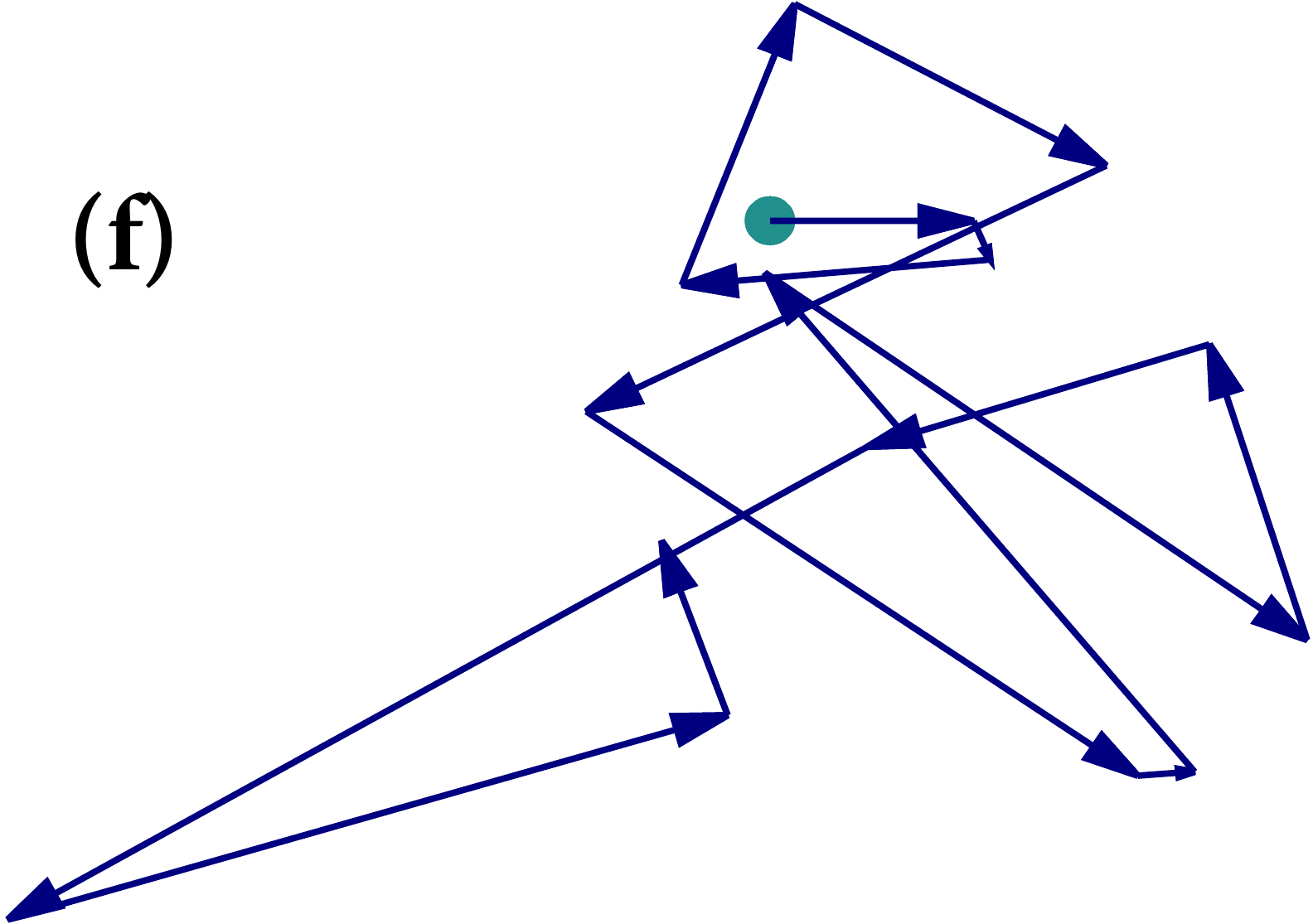}
 \caption{Schematic representation of the three different RTP dynamics considered  (Upper Panel)  and the corresponding trajectories  (Lower Panel).    (a) and (d)  correspond to the 3-state model while  (b) and (e) correspond to the   4-state model and  (c) and (f) correspond to the continuous model.}
 \label{fig:schem_all}
 \end{figure*}
Figure \ref{fig:schem_all}  (upper panel)  shows schematic representations of all the three dynamics. A set of corresponding typical trajectories are shown in Fig.~\ref{fig:schem_all}  (lower panel).

Our main goal is to investigate the behaviour of the time-dependent position probability distribution. To this end, it is convenient to recast Eq.~\eqref{eq:dynamics} as

\begin{subequations}
\bea
 \dot x (t)  &=& v_0 \sigma_x (t)  \label{eq:x-sigx} \\
\dot y (t)  &=& v_0 \sigma_y (t)  \label{eq:y-sigy},
\eea
\end{subequations}
where  $\sigma_x=\cos \theta$ and $\sigma_y= \sin \theta.$  The above equations are reminiscent of a $2$D Brownian particle  where $\sigma_x$ and $\sigma_y$ play the role of the noise. For the RTP dynamics, these effective noises, however, are very different than the delta-correlated white noise which appears in the passive Brownian case. In all the cases considered here, the auto-correlation of the effective noise has an exponential form 
\bea
\la \sigma_x (s)  \sigma_x (s')  \ra \sim e^{-a_0 \gamma\mid s-s'\mid}
\eea
and similarly for $\sigma_y$. $a_0$ is some numerical constant depending on the specific dynamics of the model. For any finite $\gamma$, the correlation decays exponentially which means that the noise is strongly correlated at short times ($|s-s'| \ll \gamma^{-1}$). It may be noted that in the limit of $\gamma\rightarrow \infty $, $\sigma$ approaches a $\delta$-correlated white noise.

This exponential nature of the auto-correlation of the effective noise
is a typical feature of the active particle dynamics and gives rise to strong memory effects in the short-time regime \cite{ABP2018}. 
In particular, we expect signatures of activity in the short-time regime. We will show below that the short-time dynamics depends crucially on the microscopic dynamics, in certain cases also giving rise to anisotropy. On the other hand, at long-times a typical Gaussian behaviour is expected. However, the signature of activity is still expected to remain in the atypical fluctuations of the position.

\begin{figure*}[t]
 \centering
 \includegraphics[width=17cm]{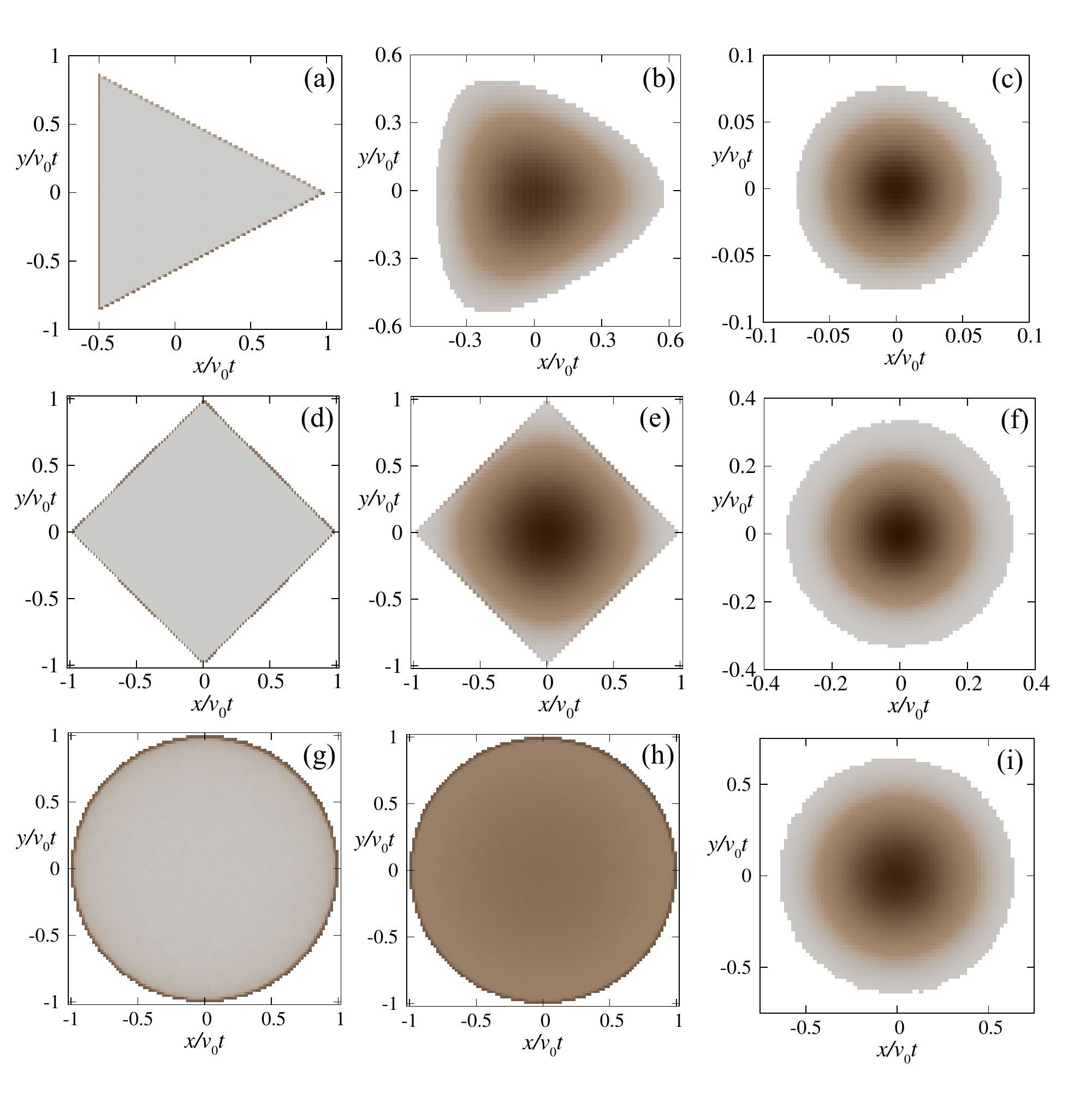}
 \caption{Plot of the two-dimensional position distribution $P (x,y;t) $, obtained from numerical simulations, of $n=3$  ( (a) ,  (b) ,  (c) ) ; $n=4$  ( (d) ,  (e) ,  (f) ) ; and continuous  ( (g) ,  (h) ,  (i) )  models for $\gamma=0.1$ The left, middle and right panels correspond to $t\ll \gamma^{-1}$  ( (a) ,  (d) ,  (g) ) ; $t\sim\gamma^{-1}$  ( (b) ,  (e) ,  (h) ) ; and $t\gg\gamma^{-1}$  (c, f, i)  respectively. The lighter grey shade indicates lower values of $P (x,y;t) $ while darker shades indicate progressively higher values. Here we have used $v_0=1.$} \label{fig:2dcont}
\end{figure*}

The change in nature of the motion of these 2D run-and-tumble particles is illustrated in Fig.~\ref{fig:2dcont} where we show  the  time evolution of the position probability distribution in the $x-y$ plane, obtained from numerical simulations. The left most panel corresponds to a time $t \ll \gamma^{-1}.$ Clearly, in this regime the shape of the probability distribution is very different in all the three models. However, there is one common feature, namely, the distribution attains its maximum value along some curve which is away from the origin implying the particle is likely to be away from the origin. This feature is similar to what has been observed in other active particle models, like active Brownian Particles etc \cite{RTP_free, ABP2018}. As time increases, the distribution changes its shape, the peak shifts towards the origin, and at long times $t \gg \gamma^{-1}$ a single-peaked Gaussian-like distribution is observed.

In this paper we present an analytical understanding of these dynamical features by investigating the position probability distribution. Here we present a brief summary of our results. 

\begin{itemize}
 \item We show that at short-time regime, the RTP shows a ballistic behaviour, \ie, in this regime, the mean-squared displacement $\propto v_{\text{eff}}^2 t^2$ where the effective velocity depends on the specific model. On the other hand, in the long-time regime, the RTP shows a diffusive behaviour, \ie, the mean-squared displacement grows linearly with time $\sim 2 D_{\text{eff}} t,$ where, the effective diffusion constant $D_{\text{eff}}$ is also model specific.
 
 \item  The symmetry of the internal spin dynamics manifests in the short-time behaviour of the probability distribution (see Fig.~\ref{fig:2dcont}). The particles cluster away from the origin along some boundary whose shape depends crucially on the microscopic dynamics. These features disappear at long times, where the crowding is near the origin.
 
 \item We calculate the time-dependent marginal position distributions, and also the full two-dimensional distribution for the continuous case. We show that the ballistic to diffusive crossover is associated to qualitatively different behaviours of the marginal position distributions at the short-time and long-time regimes. We characterize these by obtaining closed form expressions for the distributions at the two regimes.
 
 \item Investigation of the behaviour of the probability distribution functions in the long-time regime shows that, independent of the model, the typical position fluctuations are Gaussian. However, the atypical fluctuations are different for the different models and are characterized by large deviation function which we calculate explicitly. 
 
\end{itemize}
In the following three sections we study in detail the three models described above.

\section{Three-state  ($n=3$)  dynamics}\label{sec:3st}

In this Section we consider the case where the internal `spin' or the orientational degree can take three discrete values $\theta=0,2\pi/3,4\pi/3.$ The orientation changes by a rotation of $\pm 2\pi/3$   (clockwise or anti-clockwise)  with rate $\gamma/2$  (see Fig.~\ref{fig:schem_all} (a)  for a schematic representation). 
The position of the particle $ (x (t) ,y (t) ) $ evolves according to the Langevin equation Eq.~\eqref{eq:dynamics}. A typical trajectory of the particle, starting from the origin and oriented along $\theta=0$ is shown in Fig.~\ref{fig:schem_all}(d).

The time evolution of the corresponding $2$-dimensional position distribution $P(x,y,t)$ obtained from numerical simulations is shown in Fig.~\ref{fig:2dcont}(upper panel). At short-times $t \ll \gamma^{-1}$ we see a crowding away from the centre, along the boundary of a triangular region (see Fig.~\ref{fig:2dcont}(a)). To understand this behaviour, let us first note that, starting from the origin, the particle can cover a maximum distance of $v_0 t$ along its initial orientation, if there are no flips during this interval $[0,t].$
For the three different values of the initial $\theta_0$ these corresponds to the points $(v_0 t,0),~(-\frac{v_0 t}{2},\frac{\sqrt{3}v_0 t}{2})$ and $(-\frac{v_0 t}{2},-\frac{\sqrt{3}v_0 t}{2})$ in the $x-y$ plane. For one or more flips, even though the total length traversed by the particle remains $v_0t,$ the net distance covered is smaller. Thus, the position of the particle, at any time $t,$ is always bounded by the triangle formed by the above three points.  
It should be noted that this boundary can be reached by directed paths only, \ie ~say the side of the triangle between $(v_0 t,0)$ and $(-\frac{v_0 t}{2},\frac{\sqrt{3}v_0 t}{2})$ is formed by particles which start with $\theta=0$ or $\frac{2\pi}{3}$ and till time $t$, flip in between these two states only, while flip to any other state, \ie ~$\theta=\frac{4\pi}{3}$ here, would result in some point inside of the said boundary. Similarly the other two sides of the triangle can be explained. As time increases the probability of such directed paths decrease and the centre starts populating. As is evident from Fig.~\ref{fig:2dcont}(b) and (c), the population at the centre increases and we get a centrally peaked distribution at times larger than $\gamma^{-1}$.

We are interested in the position probability distribution $P (x,y,t)  = \sum_{\theta} \cal P_\theta (x,y,t) $ where   $\cal P_\theta (x,y,t) $ denotes the probability that at time $t$ the RTP has a position $ (x,y) $ and orientation $\theta.$  It is straightforward to write the corresponding Fokker-Planck  (FP)  equations,
\bea
\frac{\partial}{\partial t} \cal P_0 &=& -v_0 \frac{\partial \cal P_0}{\partial x}   + \frac \gamma2   (\cal P_{\frac{2\pi} 3}+\cal P_{\frac{4 \pi}3}) - \gamma \cal P_0   \cr
\frac{\partial}{\partial t} \cal P_{\frac{2\pi}3} &=& \frac{v_0}2  \frac{\partial \cal P_{\frac{2\pi}3}}{\partial x}  -\frac{v_0\sqrt{3}}2 
\frac{\partial \cal P_{\frac{2\pi}3}}{\partial y} + \frac \gamma2   (\cal P_{0}+\cal P_{\frac{4 \pi}3}) - \gamma \cal P_{\frac{2\pi}3} \cr
\frac{\partial}{\partial t} \cal P_{\frac{4\pi}3} &=& \frac{v_0}2  \frac{\partial \cal P_{\frac{4\pi}3}}{\partial x}  +\frac{v_0\sqrt{3}}2 
\frac{\partial \cal P_{\frac{4\pi}3}}{\partial y} + \frac \gamma2   (\cal P_{0}+\cal P_{\frac{2 \pi}3}) - \gamma \cal P_{\frac{4\pi}3}.\cr
&&\label{eq:FP_3stxy}
\eea
Here we have suppressed the argument of $\cal P_\theta$ for brevity. 
It is possible to formally solve these coupled first order differential equations using Fourier transformation. However, it is hard to invert the Fourier transformation to extract information about the spatial position distribution. Instead, in the following, we look at the evolution of the $x$ and $y$-components separately and calculate the marginal distributions which, with a slight abuse of notation we denote as $P (x,t) $ and $P (y,t)$ for simplicity.

\subsection{Marginal distribution along $x$-axis}\label{sec:3st_x}

The $x$-component of the position of the 3-state RTP evolves following Eq.~\eqref{eq:x-sigx}. Hence,  starting from the origin $x=0$ at time $t=0,$ the position at time $t$ is given by,
\bea
x (t)  = v_0 \int_0^t \id s ~\sigma_x  (s).
\eea
Here $\sigma_x = \cos \theta$ takes two distinct values $1,-\frac 12.$
Note that, at any time $t,$ $x (t) $ is bounded in the region $-v_0 t/2 \le x (t)  \le v_0 t.$

\begin{figure}[t]
 \includegraphics[width=5 cm]{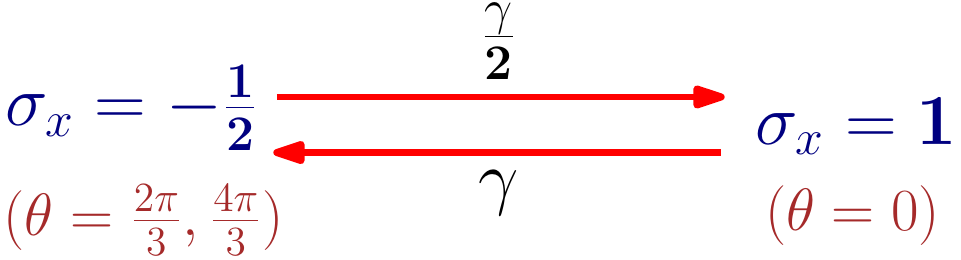}
 \caption{The effective 2-state jump process characterizing the time-evolution of $\sigma_x$ for the 3-state model.} \label{fig:sigx_3st}
\end{figure}

To understand the nature of the marginal position distribution $P (x,t) $ let us first look at the dynamical behaviour of the effective noise $\sigma_x.$  $\sigma_x$ can jump from $1$ to $-\frac 12$ through two channels, namely, $(\theta =0) \to (\theta = 2\pi/3)$ and  $(\theta =0) \to (\theta = 4\pi/3)$ and hence the jump rate for $\sigma_x=1 \to -\frac 12$ is given by $\gamma.$ On the other hand the jump, $\sigma_x=-\frac 12 \to 1$ corresponds to either $(\theta=2\pi /3) \to (\theta=0)$ or $(\theta=4\pi /3) \to (\theta=0)$ and the corresponding jump rate is just $\gamma/2.$ This effective dynamics is shown schematically in Fig.~\ref{fig:sigx_3st}. Note that we consider a uniform initial condition for $\theta$ and hence the $\sigma_x$ process is stationary at all times $t$ with $\la \sigma_x (t) \ra=0.$  It is instructive to calculate the auto-correlation of $\sigma_x$ (see Appendix ~\ref{sec:app_theta}), 
\bea
\la \sigma_x (s)  \sigma_x (s')  \ra = \frac 12 \exp \left[-\frac 32 \gamma |s-s'| \right]. \label{eq:3st_sigx_var}
\eea
As already mentioned in the previous section, the exponential form of the auto-correlator indicates that the noise is highly correlated at the short-time regime and consequently one can expect strong signatures of activity in this regime.

The simplest way to see these signatures is to look at the behaviour of the moments. As a direct consequence of the fact that $\la \sigma_x (t)  \ra=0,$ the mean position $\la x (t)  \ra$ vanishes at all times. The first non-trivial moment is then the variance $\la x^2 (t)  \ra$ which can be calculated exactly using Eq.~\eqref{eq:3st_sigx_var} and is given by,
\bea
\la x^2 (t)  \ra = \frac{2 v_0^2}{3 \gamma}\left[t - \frac 2{3 \gamma} \left(1-e^{-\frac{3\gamma t}2}\right)  \right].
\label{3stxvariance}
\eea

At short-times $t \ll \gamma^{-1}$ the variance grows quadratically,
\bea
\la x^2 (t)  \ra = v_0^2 t^2 + O (t^3) 
\eea
indicating a ballistic behaviour. Note that, the speed of the particle in this ballistic regime is simply $v_0,$ it does not depend on $\gamma.$ On the other hand, 
in the long-time regime a diffusive behaviour is recovered
\bea
\la x^2 (t)  \ra \simeq 2 D_\text{eff}\, t
\eea
where the effective diffusion constant $D_\text{eff}= v_0^2/3\gamma.$

To understand the change in behaviour in more details we investigate the position probability $P (x,t) = P_+ (x,t)  + P_- (x,t) $ where $P_+ (x,t) $  (respectively $P_- (x,t) $)  denotes the probability that position is $x$ and $\sigma_x=1$  (respectively $\sigma_x=-\frac 12$)  at time $t.$
The corresponding Fokker-Planck equations are given by
\bea
\frac{\partial P_+}{\partial t} &=& - v_0 \frac{\partial P_+}{\partial x} - \gamma P_+ + \frac \gamma 2 P_- \cr
\frac{\partial P_-}{\partial t} &=& \frac{v_0}2 \frac{\partial P_-}{\partial x} - \frac \gamma 2 P_- +  \gamma  P_+. \label{eq:FP_3st_x}
\eea
Note that this set of FP equations can also be obtained directly from Eq.~\eqref{eq:FP_3stxy} by integrating both sides over $y$ and then identifying $P_+ (x,t)  = \int d y \cal P_0 (x,y,t) $ and $P_- (x,t)  = \int d y [\cal P_{2\pi/3} (x,y,t)  + \cal P_{4\pi/3} (x,y,t) ].$ \\
We choose the initial conditions to be such that at $t=0$ the RTP can be in any of the $\sigma-$states with equal probability, \ie ,
\bea
P_+ (x,0)  = \frac 13 \delta (x)  \text{ and } ~P_- (x,0)  = \frac 23 \delta (x). 
\eea
To do this, we introduce the Laplace transform  of $P (x,t) $ w.r.t. time, 
 \bea
 \hat P_\pm (x,s)  = \int_0^\infty d t ~e^{-st} P_\pm (x,t). 
 \eea
 In terms of $\hat P (x,s) $ Eq.~\eqref{eq:FP_3st_x} reduces to,
 \bea
 v_0 \hat P_+^\prime &=& - (s+\gamma)  \hat P_+ + \frac \gamma 2 \hat P_- + \frac 13 \delta (x) \cr
 v_0 \hat P_-^\prime  &=&  (2s + \gamma) \hat P_- - 2 \gamma  \hat P_+ - \frac 43 \delta (x), 
 \label{EQ:3stxlaplace}
 \eea
 where $^\prime$ denotes the derivative with respect to $x.$ Note that the boundary condition for these equations are simply $\lim_{x \to \pm \infty} \hat P (x,s) =0.$

\begin{figure*}[t]
 \centering
 \includegraphics[width=8 cm]{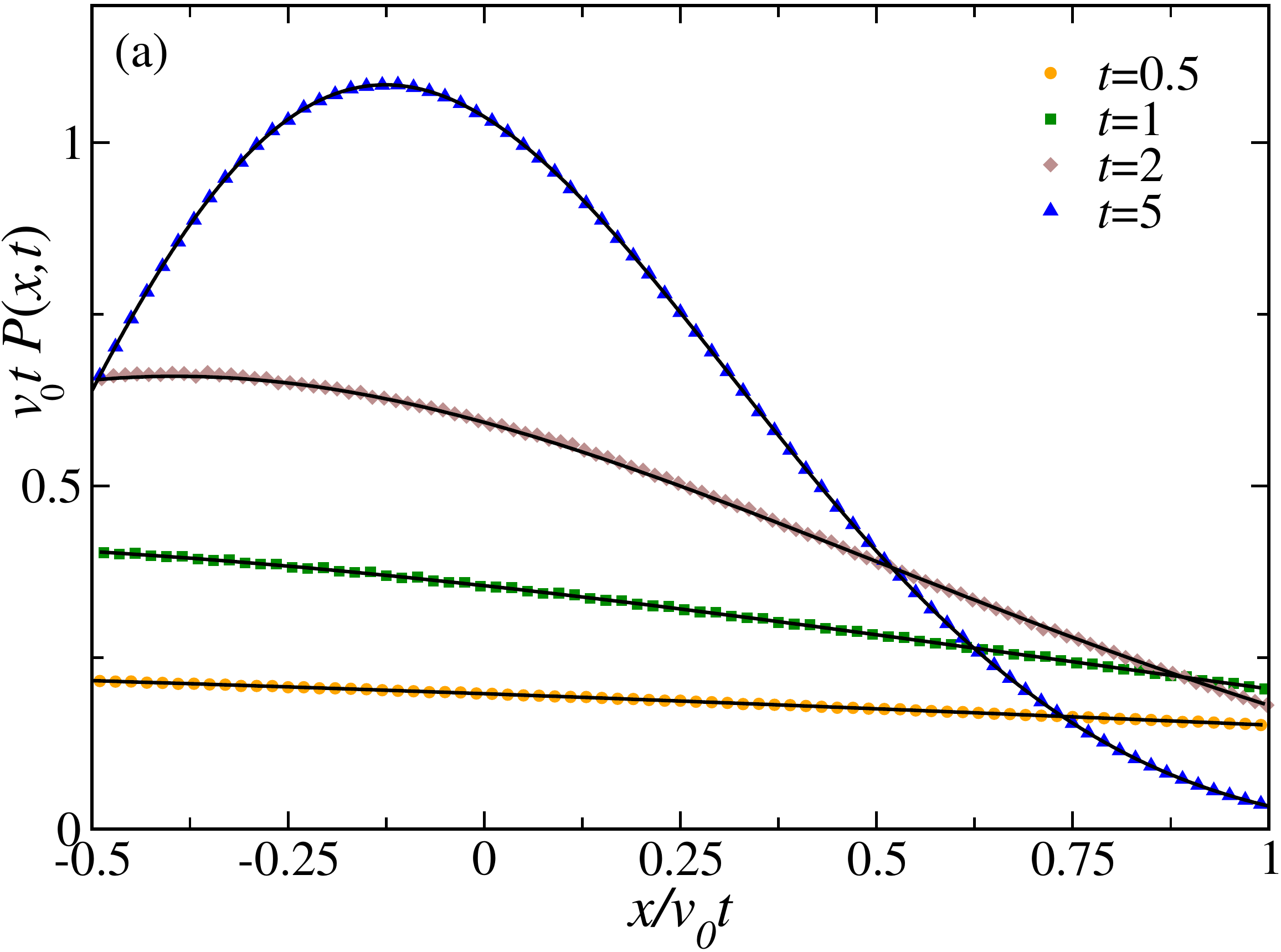}~~~~~
 \includegraphics[width=8 cm]{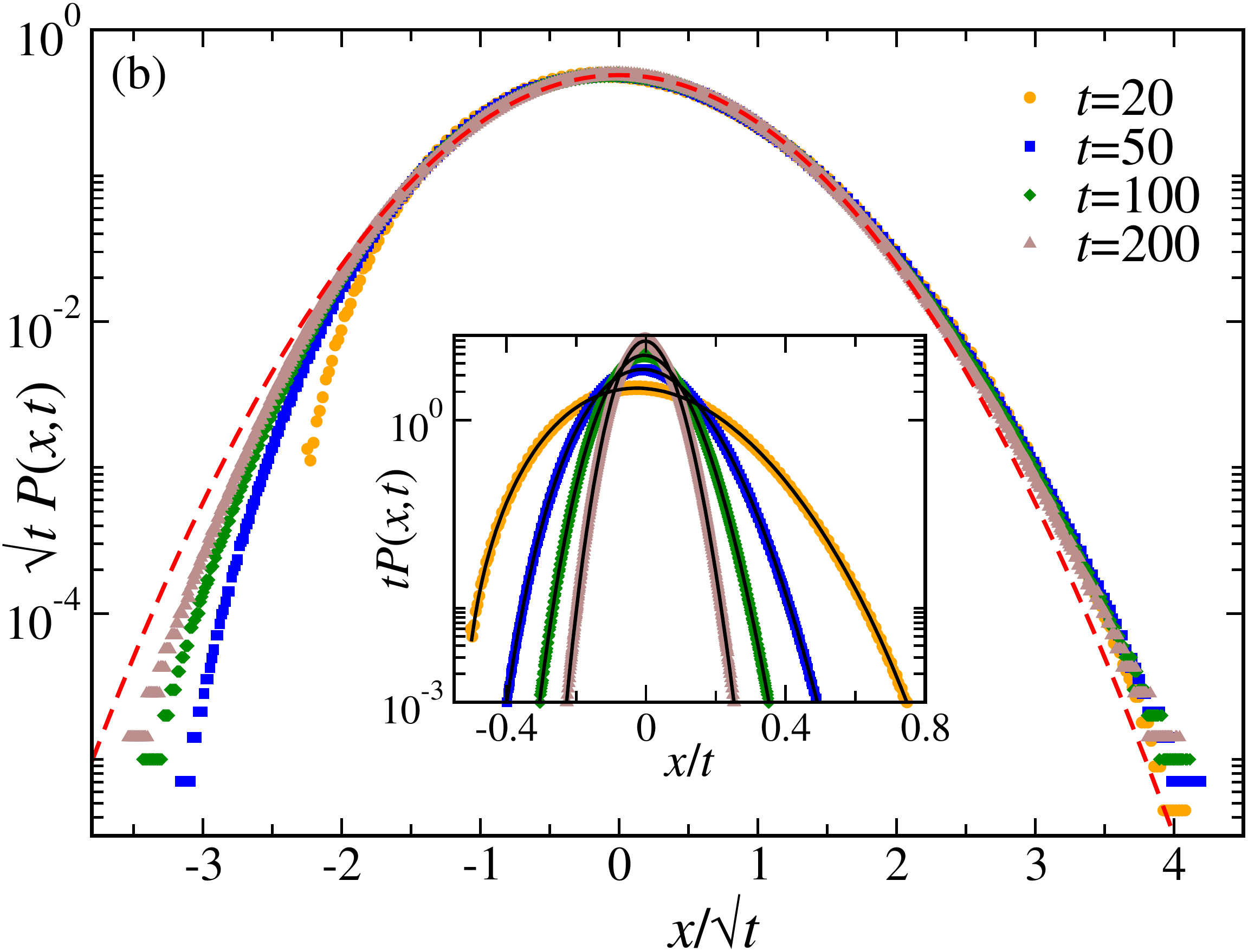}
 \caption{3-state model: (a) Plot of $P (x,t) $ for $\gamma=1$ and for different values of $t.$ The solid black lines correspond to the analytical prediction Eq.~\eqref{eq:3st_Pxt} and the symbols correspond to the data obtained from the numerical simulations. For better visibility we have excluded the delta functions at the two boundaries.
(b) Plot of $P(x,t)$ obtained from numerical simulations, as a function of the scaled variable $w=x/\sqrt{t}$ for different (large) values of $t$ and $\gamma=1.$ The red dashed line shows the Gaussian distribution (see Eq.~\eqref{eq:3st_Pw}). The inset shows the same data as a function of $x/t.$  The solid black lines there correspond to the analytical prediction Eq.~\eqref{eq:3st_Pxt}.  Here we have used $v_0=1.$ 
 }
 \label{fig:3st_px}
\end{figure*} 
 The solution of this set of coupled differential equation is obtained for $x\neq 0$ as (see Appendix \ref{sec:app_int} for details),
  \bea
\hat P (x,s)  = \left \{ 
\begin{split}
\frac{6 \gamma +5 s- \lambda}{2 v_0 \lambda} \exp \left[-\frac x{2v_0} (\lambda-s) \right] & ~ \text{for}~ x>0 \cr
\frac{6 \gamma +5 s +\lambda}{2 v_0 \lambda}~ \exp \left[\frac x{2v_0} (\lambda+s) \right] & ~ \text{for}~ x<0,
\end{split}
\right.
\label{EQ:3statexlaplace}
\eea
where where $\lambda=\sqrt{3s (3s+4\gamma) }$ .\\
To obtain $P (x,t) $ we need to invert the Laplace transformation by evaluating the Bromwich integral,
\bea
P (x,t)  = \frac 1 {2\pi i} \int_{c-i \infty}^{c+i \infty} d s ~e^{st} \hat P (x,s),
\eea
where $c$ is a real number chosen such that all the singularities of the integrand lies on the left side of the vertical contour from $c-i\infty$ to $c+i\infty$ on the complex $s$ plane.
This integral, which involves a Branch-cut along the negative real $s$-axis, can be computed  as detailed in Appendix \ref{sec:app_int}. Finally, we have,
\bea
P (x,t)  &=& \Theta (v_0 t-x) \Theta \left(x+\frac{v_0 t}2\right)  \, G_x (x,t)  \cr
&+& \frac {1}3 e^{-\gamma t}\delta (x- v_0 t)  + \frac {2}3 e^{-\frac{\gamma t}2} \delta \left(x+ \frac{v_0 t}2\right).  
\label{eq:3st_Pxt}
\eea
Here $G_x (x,t) $ is the bulk distribution, obtained from the branch-cut integral, whose explicit form is given below in Eq.~\eqref{eq:3st_Gxt}.  The Dirac-delta functions at $x=v_0 t$ and $x=-v_0t/2$ correspond to the cases where initially $\sigma_x=1$  (respectively $-\frac 12$)  and $\sigma_x$ does not change its value up to time $t.$ Presence of such delta-functions are typical to RTP-like dynamics in free space, and has been observed also for one-dimensional RTP \cite{RTP_free}. The presence of the $\Theta$-functions multiplying $G_x(x,t)$ alludes to the fact that, at any time $t,$ the particle is always bounded between $x=v_0 t$ and $x=-v_0t/2.$

The bulk distribution $G_x (x,t)$, obtained from the branch cut integral is (see Appendix \ref{sec:app_int}),
\bea
G_x(x,t)  &=& \frac 1{6 \pi v_0}\int_0^{a} d u~e^{-u (t+\frac{x}{2v_0}) }
\left[- 3\sin \frac{3x}{2 v_0}\sqrt{u (a-u) } \right. \cr
&& \left. + \frac{ (6\gamma - 5u) }{\sqrt{u (a-u) }}\cos \frac{3x}{2 v_0}\sqrt{u (a-u)}  \right], \label{eq:3st_Gxt}
\eea
where $a=4\gamma/3$. 
 Upon doing this integral (See Appendix \ref{sec:app_int} for details), we get,
\begin{align}
&G_x(x=zv_0t, t)= \frac{\gamma e^{-\frac{\gamma t}{3} (z+2)}}{9 v_0} \Biggl[4 I_0 \left(\frac{2\, \gamma t }{3} \sqrt{(2z+1) (1-z)}\right) \notag\\
&+ \frac{5-2z}{\sqrt{(2z+1) (1-z)}}I_1 \left(\frac{2\, \gamma t }{3} \sqrt{(2z+1) (1-z)}\right) 
\Biggr],
\label{eq:Gx-exact}
\end{align}
where $I_\nu(z)$ is the modified Bessel function of the first kind~\cite{dlmf}. \\
Figure~\ref{fig:3st_px} compares the exact analytical $P (x,t) $   for different values of $t$  with the same obtained from numerical simulations.

As mentioned already, we are particularly interested in the behaviour of $P (x,t) $ in the short-time  ($t \ll \gamma^{-1}$)  and long-time  ($t \gg \gamma^{-1}$)  regimes. At short times, Taylor expanding the right hand side of Eq.~\eqref{eq:Gx-exact} around $t=0,$  we get,
\begin{equation}
G_x (x=zv_0 t, t) = \frac {2 \gamma}{9 v_0}\Bigl[ 2- \bigg(z+\frac 12\bigg) \gamma t  + O(t^2) \Bigr].
\end{equation}
Clearly, the distribution is linear in the bulk while the $\delta$-function dominates at the boundaries. This linear nature of $P (x,t) $ at short times is clearly visible from the $t=0.5$ curve in Fig.~\ref{fig:3st_px} a.

At long times  ($t \gg \gamma^{-1}$), using the asymptotic behavior of Bessel functions, we have the large deviation form
\begin{equation}
P(x=z v_0 t,t) \sim e^{-t \phi(z)},
\end{equation}
where the large deviation function is given by
\begin{equation}
\phi(z)=\frac{\gamma}{3} \Bigl[  z+2 -2\sqrt{(2z+1)(1-z)} \Bigr]. 
\end{equation}
Around $z=0,$ the large deviation function is quadratic,
\bea
\phi(z) = \frac 34 \gamma z^2 + O(z^3).
\eea
Consequently, the typical fluctuations of $x$ around the origin are of the order $\sqrt{t}$ and are Gaussian in nature, \ie, the distribution of the scaled variable $w=x/\sqrt{t}$ is given by 
\begin{equation}
P (w,t)  \simeq \sqrt{\frac{3 \gamma}{4 \pi v_0^2}}\, \exp{\left(- \frac{3 \gamma w^2}{4v_0^2}\right)}. \label{eq:3st_Pw}
\end{equation}
Figure \ref{fig:3st_px}(b) shows a plot of $P(w,t)$ as a function of the scaled variable $w = x/\sqrt{t}$ which leads to a scaling collapse following Eq.~\eqref{eq:3st_Pw} near the peak at $w=0.$ However, the signature of the active nature of the system is clearly visible at the tails where the distribution remains non-Gaussian.

\begin{figure}[t]
\includegraphics[width=8 cm]{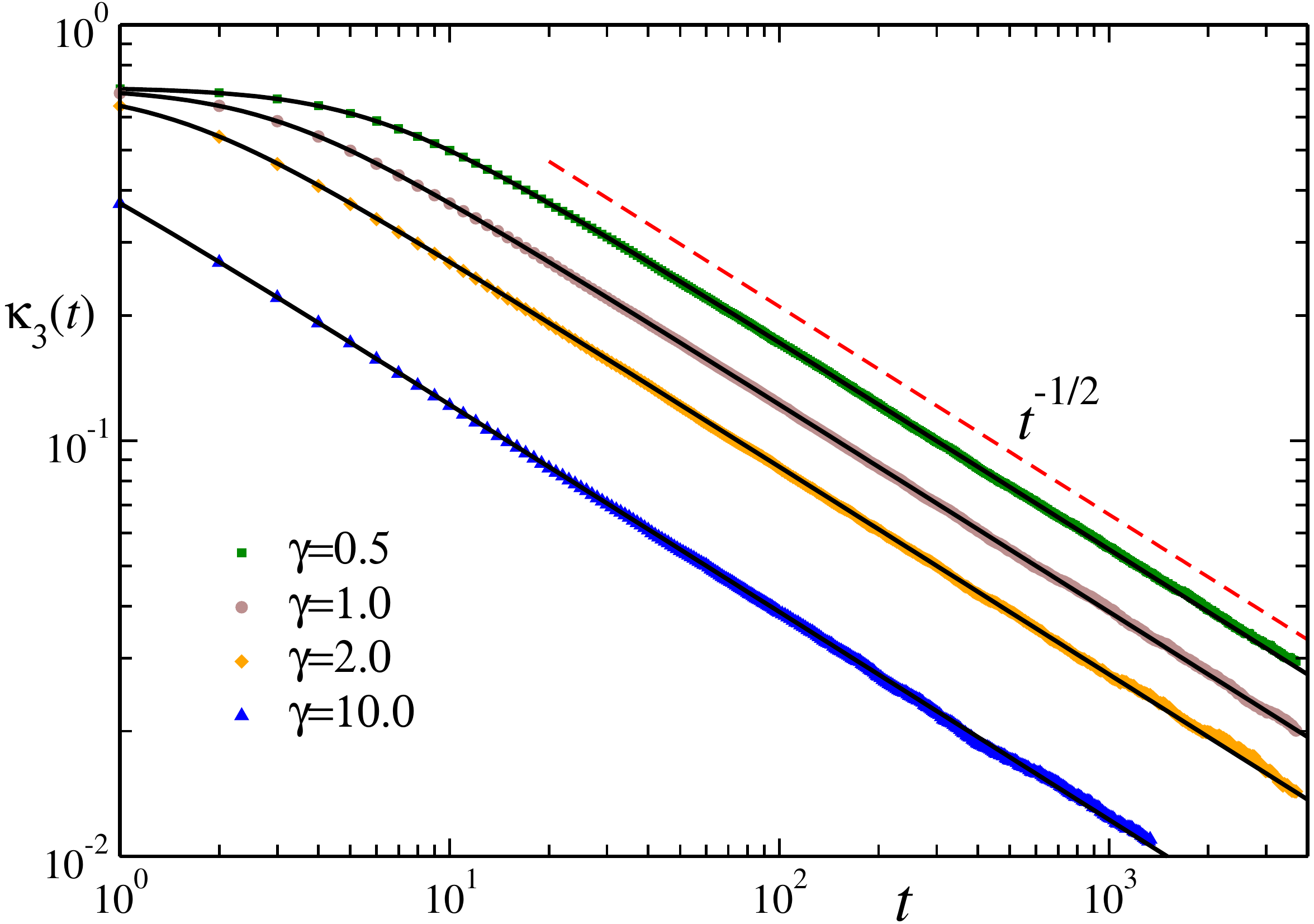}
\caption{Time evolution of the skewness of the $x$-marginal distribution for the $3$-state process for different values of $\gamma.$ The solid black lines indicate the analytical prediction and the symbols show the data from numerical simulations.}\label{3stxSkew}
\end{figure}

Another interesting feature of $P(x,t)$ is that it is asymmetric about the origin.  To quantify the asymmetry we calculate the skewness  which is defined in terms of second and third cumulants. In this case the first moment $\langle x\rangle=0$, and hence, the skewness is given by, 
\bea
\kappa_3&=&\frac{\langle x^3\rangle}{\langle x^2\rangle^{3/2}}.
\eea
To calculate the third moment, we need the three point $\sigma-$ correlation, which can be calculated using Eq.~\eqref{eq:th_prop} and turns  out to be (see Appendix \ref{sec:app_theta})
\bea
\langle\sigma_x (s_1) \sigma_x (s_2) \sigma_x (s_3) \rangle &=&\frac 14 e^{-\frac{3}{2}\gamma |s_3-s_1|}
\eea
Thus, the third moment is given by,
\bea
\langle x^3(t)\rangle &=&\frac{2 v_0^3}{9\gamma^3}\left ( (4+3\gamma t) e^{-\frac{3\gamma t}{2}} + 3\gamma t -4\right). 
\eea
Using the above expression and Eq.~\eqref{3stxvariance}, $\kappa_3$ can be easily calculated. It turns out that $\kappa_3(t)>0$ for all finite $t$ indicating a positively skewed distribution $P(x,t)$. Fig.~\ref{3stxSkew} shows a plot $\kappa_3 (t) $ as a function of time $t.$ At large times, $\kappa_3 (t) $ decays algebraically,
\bea
\kappa_3 (t) \sim\sqrt{\frac{3}{2\gamma t}}~~~~\text{as}~~t\to \infty,
\eea
indicating a very slow approach towards a symmetric distribution.

\subsection{Marginal distribution along $y$-axis}\label{sec:3st_y}

A direct consequence of the inherent anisotropy of the $3$-state model is that the time evolution of the $y-$component of position is very different from its $x$ counterpart. In this section we focus on the marginal distribution $P (y,t) $ of the 3-state model. 

\begin{figure}[t]
\includegraphics[width=5 cm]{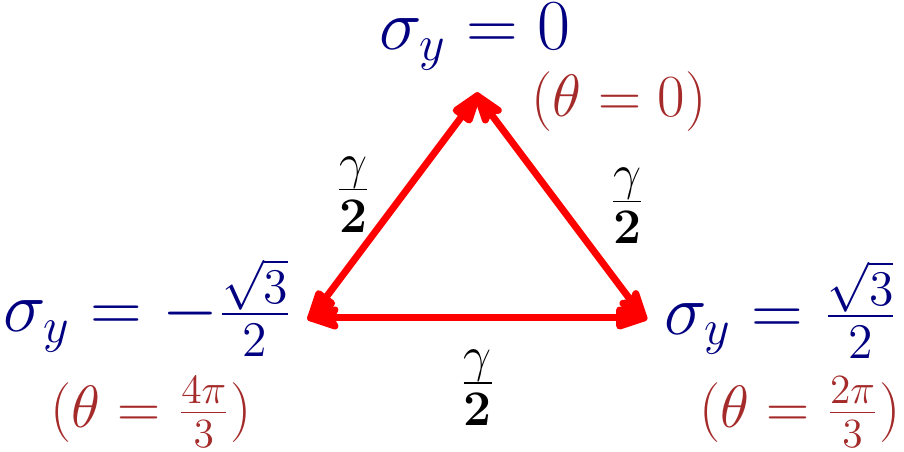}
\caption{3-state model: Schematic representation of the jump-process governing the time evolution of $\sigma_y.$}\label{fig:schem_3sty}
\end{figure}

In this case, the effective equation of motion along $y$ is given by Eq.~\eqref{eq:y-sigy}. Consequently, starting from the origin $y=0$ at time $t=0,$ we have,
\bea
y (t) =v_0 \int_0^t d s~ \sigma_y (s)  \label{eq:yt_3st}
\eea
where $\sigma_y=\sin\theta$ takes $3$ distinct values $0, \pm \sqrt{3}/2.$ Thus, $\sigma_y$ evolves according to a $3$-state jump process, with the jump rates being $\frac{\gamma}{2}$ for all the transitions  (see Fig.~~\ref{fig:schem_3sty} for a schematic representation of the $\sigma_y$ process). Note that, at any time $t,$ $y (t) $ is bounded between $[-\frac{\sqrt{3}}{2}v_0 t,\frac{\sqrt{3}}{2}v_0 t].$

As before, we first look at the moments to get an idea about the behaviour of this effective $1-$d process. Similar to the $x$-component, the first moment $\la y (t)  \ra$ vanishes at all times. The second moment can be calculated in a straight-forward manner using the auto-correlation of $\sigma_y (t) ,$ which is same as that of $\sigma_x (t) $
\bea
\la \sigma_y (s) \sigma_y (s') \ra  = \frac 12 \exp \left[-\frac 32 \gamma |s-s'| \right] \label{eq:3st_sigy_var}
\eea
Consequently, the variance, 
\bea
\la y^2 (t)  \ra = \frac{2 v_0^2}{3 \gamma}\left[t - \frac 2{3 \gamma}\bigg  (1-e^{-3\gamma t/2} \bigg)  \right]
\eea
is identical with $ \la x^2 (t)  \ra.$

 Hence, once again we see a ballistic behaviour at short times  ($t\ll \gamma^{-1}$) , $\la y^2\ra\sim v_0^2 t^2,$ which goes over to a long-time diffusive behaviour ($t\gg \gamma^{-1}$)  with $\la   y^2\ra \sim \frac{2v_0^2}{3\gamma} t$. So the effective diffusion constant $D_{\text{eff}}=\frac{v_0^2}{3\gamma}$, same as for the $x$ component. Though the qualitative short and long time behaviours are similar, the $x$ and $y$ motions are very different which is evident from the Fig.~\ref{fig:2dcont} (a), (b), and will become more clear from the full distribution $P (y,t) $ which we study below.

To calculate the time-dependent distribution $P (y,t) $ of the $y$-component, we proceed in the same way as before and write the FP equations for $P_{\alpha} (y,t) ,$  which denotes probability of finding the particle at position $y$ at time $t$ with $\sigma_y= \alpha.$ Note that for notational simplicity we denote the marginal probability distribution of the $y-$component also with the letter $P$.
The corresponding Fokker-Planck equations are,
\bea
\frac{\partial P_0}{\partial t}&=&-\gamma P_0+\frac{\gamma}{2} (P_+ + P_-) \nonumber\\
\frac{\partial P_+}{\partial t}&=&-v\frac{\partial P_+}{\partial y}-\gamma P_++\frac{\gamma}{2} (P_0 + P_-) \nonumber\\
\frac{\partial P_-}{\partial t}&=&v\frac{\partial P_-}{\partial y}-\gamma P_- +\frac{\gamma}{2} (P_0 + P_+) 
\label{eq:3stateYFP}
\eea
 where we have denoted $v=\frac{\sqrt{3}}{2}v_0;$  we have suppressed the argument of the $P_\alpha$ in the above equation for brevity. The initial conditions are chosen in such a way that all the three values of $\sigma_y$ are equally likely at time $t=0$ and  since we consider that the particle starts from the origin, we must have
 \bea
 P_\alpha (y,0)  = \frac 13 \delta (y)  \quad \forall \alpha.
 \eea
 
 We follow the same procedure as in the previous section and introduce a Laplace transformation w.r.t. time $t,$ 
 \bea
 \hat P_\alpha (y,s)  = \int_0^{\infty}dt~e^{-st}P_\alpha (y,t).  
 \eea
 Upon doing the Laplace transform, Eqs.~\eqref{eq:3stateYFP} become
 \bea
 \hat{P}_0 &=&\frac{\gamma}{2(s+\gamma)}(\hat{P}_+ +\hat{P}_-)+\frac{\delta(y)}{3(s+\gamma)},\nonumber\\
 v\hat{P}'_+&=&-(s+\gamma)\hat{P}_+ + \frac{\gamma}{2}(\hat{P}_0 +\hat{P}_-)+\frac{\delta(y)}{3},\nonumber\\
 v\hat{P}'_-&=&(s+\gamma)\hat{P}_- + \frac{\gamma}{2}(\hat{P}_0 +\hat{P}_+)-\frac{\delta(y)}{3}.
 \label{eq:3st_y_p0p1p2}
 \eea
where $'$ denotes the derivative with respect to y. We want the full distribution, \ie,  $\hat{P}=\hat{P}_0 +\hat{P}_-+\hat{P}_+$.

 Solving Eqs.~\eqref{eq:3st_y_p0p1p2} we get,
 \bea
\hat P (y,s) &=&\frac{ (2s+3\gamma) ^2}{12 v\sqrt{s} (s+\gamma) ^{\frac{3}{2}}}\exp{\left[-{\frac{ (2s+3\gamma) }{2 v}\sqrt{\frac{s}{s+\gamma}}}|y| \right]} \cr
&& + \frac{\delta (y) }{3 (s+\gamma)}.  \label{eq:3st_Pys}
\eea 
To find the position distribution as a function of the time $t$  we need to compute the inverse Laplace transformation of $\hat{P} (y,s).$ Let us first note that, the last term in Eq.~\eqref{eq:3st_Pys}, when inverted, results in $\frac{1}{3}e^{-\gamma t}\delta (y) $, which denotes the probability that the particle started with $\sigma_y=0$ and $\sigma_y$ did not flip up to time $t.$ To invert the first, more complicated term  (in Eq.~\eqref{eq:3st_Pys}) , one needs to compute a Bromwich Integral in the complex $s$-plane. It is easy to see that this integral involves a Branch-cut along the negative $s$-axis which can be converted to a real line integral following the same procedure as in Sec.~\ref{sec:3st_x}  (see Appendix \ref{sec:app_int}). 
Finally, we have,
\bea
P (y,t) &=& G_y (y,t) \Theta (v t -|y|) \cr
& +& \frac{e^{-\gamma t}}{3}\bigg[\delta (y)  +\delta (y-v t) +\delta (y+v t) \bigg]
\eea
where,
\bea
G_y (y,t) &=&\int_{0}^{\gamma-\epsilon} \frac{d u~e^{-u t}}{12\pi v}\frac{ (3\gamma-2 u) ^2}{\sqrt{u} (\gamma-u) ^{\frac{3}{2}}} \cos \bigg[\frac{ (3\gamma-2 u) y}{2 v}\sqrt{\frac{u}{ (\gamma-u) }}\bigg]\cr &-&\frac{e^{-\gamma t}}{3 \pi y}\sin\left(\frac{\gamma ^{3/2}y}{2v\sqrt{\epsilon}}\right)+O(\sqrt{\epsilon}).
\label{3stateYfull}
\eea
 where $\epsilon$ is a very small number. Eq.~\eqref{3stateYfull} can be evaluated numerically for small $\epsilon$. It turns out that this agrees well with numerical simulations for times greater than $\gamma ^{-1}$. For $t\lesssim \gamma^{-1}$  numerical evaluaation of Eq.~\eqref{3stateYfull} becomes difficult. In this regime we adopt a different approach and  write $\hat P(y,s)$ in Eq.~\eqref{eq:3st_Pys}  as a series in $s$ for $y\neq 0$,
\bea
\hat P(y \ne 0,s) &=&\frac{1}{12 v}\sum_{n=0}^{\infty}\frac{(-y/v)^n}{2^n n!} (3\gamma)^{n+2}\sum_{m=0}^{n+2} {n+2\choose m}\left(\frac{2}{3\gamma}\right)^m\n\\&&\times \frac{s^{m+(n-1/2)}}{(s+\gamma)^{n+3/2}}.
\eea
Then, taking the inverse Laplace transformation of the above equation with respect to $s$ gives 
\bea
G(y=zvt,t) &=&\frac{9\gamma^2 t}{12 v}\sum_{n=0}^{\infty}\frac{(-|z|)^n}{2^n n!} (3\gamma t)^{n}\sum_{m=0}^{n+2} {n+2\choose m}\nonumber \\&\times&\left(\frac{2}{3\gamma t}\right)^m {}_1\bar{F}_1\left(\frac{n+3}{2},2-m,-\gamma t\right),
\label{eq:3st_y_series}
\eea
where ${}_p\bar{F}_q(a,b,z)$ is the regularized Hypergeometric function \cite{dlmf}.

\begin{figure*}[t]
 \centering
 \includegraphics[width=8 cm]{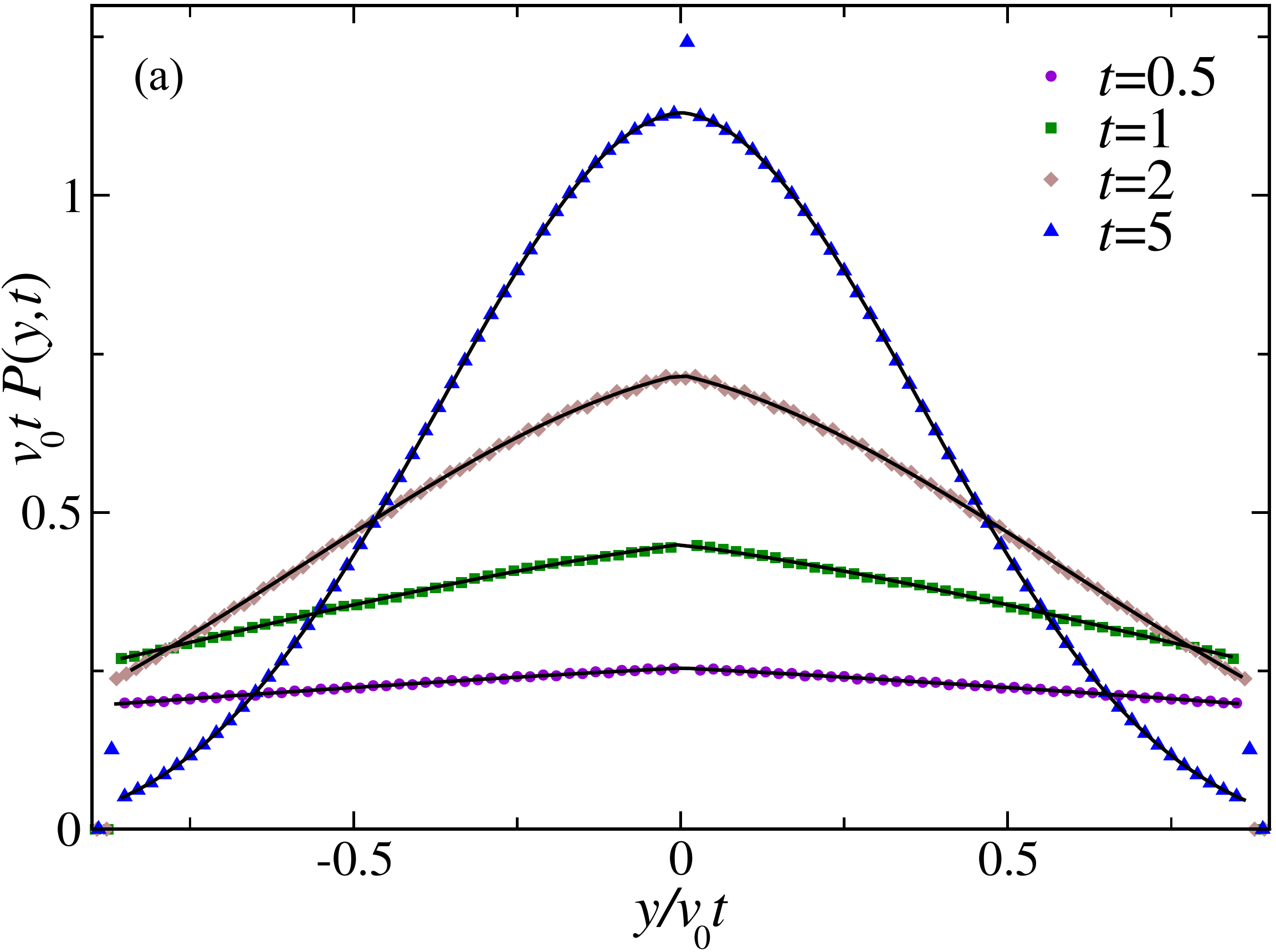}~~~~~
 \includegraphics[width=8 cm]{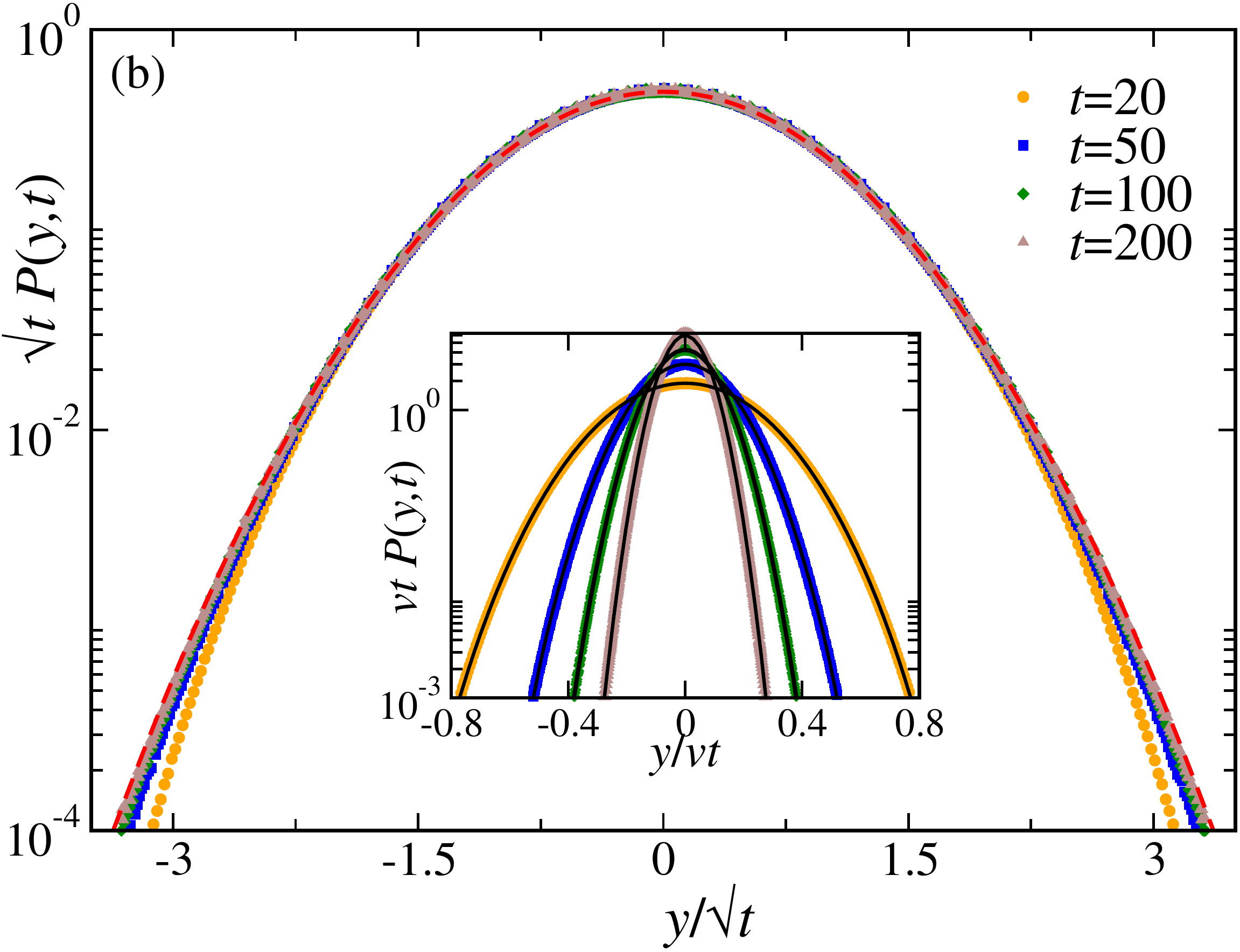}
 \caption{3-state model: (a) Plot of $P (y,t) $ for $\gamma=1$ and for different values of $t.$ The solid black lines correspond to the analytical predictions Eq.~\eqref{3styshort} for $t=0.5,1$; Eq.~\eqref{3stateYfull} for the other cases and the symbols correspond to the data obtained from the numerical simulations. For better visibility we have excluded the delta functions at the origin and the boundaries.
(b) Plot of $P(y,t)$ obtained from numerical simulations, as a function of the scaled variable $w=x/\sqrt{t}$ for different (large) values of $t$ and $\gamma=1.$ The red dashed line shows the Gaussian distribution (see Eq.~\eqref{eq:3st_y_scaled}). The inset shows the same data as a function of $x/t.$  The solid black lines there correspond to the analytical prediction Eq.~\eqref{3stateYfull}.  Here we have used $v_0=1.$ 
 }
 \label{fig:3st_py}
\end{figure*}

Using this result, we can get closed form expressions for the probability distribution function at short and long times. At short times, the distribution is dominated by the three $\delta$-functions, to get the bulk distribution it is sufficient to calculate the first few terms of the series to get the leading order behaviour. The short-time distribution thus comes out to be
\bea
G(y,t) &= &\frac{\gamma}{\sqrt{3}v_0}-\frac{\gamma^2 t}{4\sqrt{3}v_0}+\frac{\gamma^2 y}{18v_0^2}(5\gamma t-6) +O(t^2).~~
\label{3styshort}
\eea
At large times, the regularized Hypergeometric function in Eq.~\eqref{eq:3st_y_series} can be approximated to the highest order in $t,$ as
 \bea
 {}_1\bar{F}_1\left(\frac{n+3}{2},2-m,-\gamma t\right)&\approx &\frac{(\gamma t)^{(n+3)/2}}{\Gamma((1-n)/2-m)}.\nonumber
 \eea
The summation over $m$ can then be performed to give
\bea
G(y=zvt,t) &\approx &\frac{9\gamma}{12v}\sum_{n=0,2,\cdots}^{\infty}\left(-\frac{3}{2}\right)^{\frac{3n+1}{2}}\frac{ (\gamma|z|)^n}{n!}\nonumber \\&\times & \frac{U\left(\frac{1+n}{2},\frac{7+3n}{2},-\frac{3\gamma t}{2}\right)}{\Gamma\left(\frac{1-n}{2}\right)}
\eea
where $U(a,b,z)$ is the HypergeometricU function. The presence of $\Gamma((1-n)/2)$ restricts the sum to only over even $n$. Expanding the HypergeometricU function to the leading order in $t$ and using the properties of the $\Gamma$-function, we get a Gaussian in this large time regime,
 \bea 
 G(y,t) &\approx &\sqrt{\frac{3\gamma}{4\pi t v_0^2}}\exp \left(-\frac{3\gamma y^2}{4 v_0^2 t}\right).
 \label{eq:3st_y_ld}
 \eea 
As before, it is useful to introduce the scaled variable $w=y/\sqrt{t}$, which has the distribution, 
  \bea 
 P(w,t) &\approx &\sqrt{\frac{3\gamma}{4\pi  v_0^2}}\exp \left(-\frac{3\gamma w^2}{4 v_0^2 }\right).
 \label{eq:3st_y_scaled}
 \eea 
 Fig.~\ref{fig:3st_py} compares the analytical expression for the probability distribution function with numerical simulations. For times $t\lesssim \gamma^{-1}$, we use Eq.~\eqref{3styshort} while Eq.~\eqref{3stateYfull} is used for the other cases.

	Let us briefly summarize the results of the $3$-state model. We have calculated the exact time dependent marginal distributions, short time distributions for both $x$ and $y$ are dominated by $\delta-$functions, however in the bulk the leading order contribution to $x$ distribution is quadratic, while for $y$, it is linear. The $y$ distribution is symmetric at all times, unlike the $x$ distribution which is highly asymmetric at short times which decreases with time.

\section{Four state  ($n=4$)  dynamics}\label{sec:4st}

 In this section we consider the case $n=4$, \ie, where the internal spin can take 4 values, $\theta=0,\frac{\pi}{2},\pi,\frac{3\pi}{2}$. The orientation thus changes by $\pm \frac{\pi}{2}$ (\ie, clockwise or anti-clockwise)  with a rate $\frac{\gamma}{2}$  (see Fig.~\ref{fig:schem_all} (b)). A typical trajectory of the particle starting from the origin can be seen in Fig.~\ref{fig:schem_all}(e).

The time evolution of the full $2$d distribution obtained from numerical simulations is shown in Fig.~\ref{fig:2dcont}((d), (e), (f)). At time scales less than $\gamma ^{-1}$, there is a crowding away from the origin. This can be explained in the same way as the $n=3$ case, if the particle starts from the origin with $\theta=0,~\frac{\pi}{2},~\pi$ and $\frac{3\pi}{2}$ with equal probability at $t=0$, then at time $t$, it can go to $(v_0 t,0),~(0,v_0 t)$, $(-v_0 t,0)$ and $(0,-v_0 t)$ in the $x-y$ plane which  form a diamond Fig.~\ref{fig:2dcont}(f), the sides of the diamond are formed by directed paths. This marks the boundary of the distribution in the $x-y$ plane. As time increases the crowding at the boundary decreases and the centre starts populating as is evident from Fig.~\ref{fig:2dcont}(e). Finally we get a centrally peaked distribution at times larger than $\gamma^{-1}$ Fig.~\ref{fig:2dcont}(f).

This model has been introduced recently in \cite{3st-RTP2019} where the stationary distribution in the presence of external potential has been studied. Here we calculate the position distribution in the free space.
The position probability distribution $\cal P (x,y,t)  = \sum_{\theta} \cal P_\theta (x,y,t) $ where $\cal P_\theta (x,y,t) $ denotes the probability that the particle is at position $ (x,y) $ with orientation $\theta$ at time $t.$ These probabilities evolve according to the Fokker-Planck  (FP)  equations, 
\bea
\frac{\partial}{\partial t} \cal P_0 (x,y,t)  &=& - v_0\frac{\partial \cal P_0}{\partial x}  + \frac \gamma 2  (\cal P_{\frac{\pi}{2}}+\cal P_{\frac{3\pi}{2}})  - \gamma \cal P_0, \cr
\frac{\partial}{\partial t} \cal P_{\frac{\pi}{2}} (x,y,t)  &=&   -v_0 \frac{\partial \cal P_{\frac{\pi}{2}} }{\partial y} + \frac \gamma 2  (\cal P_0+\cal P_{\pi})  - \gamma \cal P_{\frac{\pi}{2}}, \cr
\frac{\partial}{\partial t} \cal P_{\pi} (x,y,t)  &=& v_0 \frac{\partial \cal  P_{\pi} }{\partial x}  + \frac \gamma 2  (\cal P_{\frac{\pi}{2}}+\cal P_{\frac{3\pi}{2}})  - \gamma \cal P_{\pi} ,\cr
\frac{\partial}{\partial t} \cal P_{\frac{3\pi}{2}} (x,y,t)  &=&  v_0 \frac{\partial \cal P_{\frac{3\pi}{2}}}{\partial y}  + \frac \gamma 2  (\cal P_0+\cal P_{\pi})  - \gamma \cal P_{\frac{3\pi}{2}}. \label{eq:FP_2d}
\eea
where the arguments of the $\cal{P}_{\theta}$s have been suppressed on the r.h.s. for brevity. These coupled differential equations again can be formally solved by writing a $4\times 4$ matrix in Fourier space, but the eigenvalues and eigenvectors are complicated and it is very hard to get the inverse transform. So as in the previous case we concentrate on the marginal distributions only.
\subsection{Marginal distribution along $x$-axis}

In this model, $\sigma_x$ and $\sigma_y $ have the same dynamics, so the process is symmetric in $x$ and $y$ at all times, unlike the $n=3$ case. Thus, it is sufficient to calculate the distribution along any one direction (say $x$).
The position $x (t) $ evolves according to the following equation,
\bea
\dot x = v_0 \sigma_x (t) 
\eea
where $\sigma_x=\cos\theta$ is the effective 3-state internal spin degree of freedom which can take 3 values, $0,\pm 1$ corresponding to $\theta=0,\frac{\pi}{2},\pi$ respectively.Here, at anytime $t$, the motion is bounded in the region $|x|<v_0 t$. 

\begin{figure}[ht]
 \includegraphics[width=5 cm]{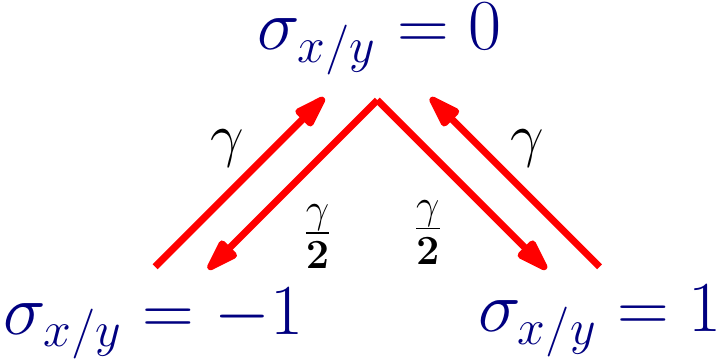}
 \caption{The effective 3-state jump process for the $4-$state model} \label{fig:4stateX}
\end{figure}

The effective noise $\sigma_x$ can jump to $0$ from $\sigma_x=\pm 1$ which corresponds to the flip in $\theta$ from $0\rightarrow\frac{\pi}{2}$ and $\pi \rightarrow\frac{\pi}{2}$. Hence the rate for these jump processes are $\gamma$ each. $\sigma_x$ can also jump from $0\rightarrow \pm 1,$ corresponding to the flips $\theta=\frac{\pi}{2}\rightarrow 0$ and $\theta=\frac{\pi}{2}\rightarrow \pi$. So the jump rates for these two processes are $\frac{\gamma}{2}$ each. This dynamics is illustrated in Fig.~\ref{fig:4stateX}.

The $\sigma_x$ process is stationary at all times with $\langle\sigma_x (t) \rangle=0$ and the autocorrelation function (see Appendix ~\ref{sec:app_theta}),
\bea
\langle \sigma_x (s) \sigma_x (s') \ra &=&\frac{1}{2}e^{-\gamma\mid s'-s\mid}.
\label{EQ:4state_sigmacorr}
\eea
Though the qualitative behaviour of the $\sigma-$correlations are very similar to the $n=3$ case, the decay constant is different. Using Eq.~\eqref{EQ:4state_sigmacorr} we can readily calculate the first two moments of $x(t)$. The mean, $\langle x (t)  \rangle$, is zero at all times as  $\langle\sigma_x (t) \rangle=0$, while using Eq.~\eqref{EQ:4state_sigmacorr} the variance comes out to be 
\bea 
\langle x(t)^2\rangle &=& \frac{v_0^2}{\gamma^2} \left [\gamma t- (1-e^{-\gamma  t}) \right]. 
\eea
So, at short times ($t\ll \gamma^{-1}$) ,
\bea
\langle  x^2(t) \rangle=\frac{v_0^2 t^2}{2}+O (t^3) .
\eea
Thus indicating ballistic behaviour with an effective speed $v_\text{eff}= v_0 / \sqrt{2}$. However, at long times  ($t\gg\gamma^{-1}$) ,
\bea
\langle x^2(t) \rangle\approx\frac{v_0^2 t}{\gamma}. \n
\eea
\ie, the motion is diffusive with  $D_{\text{eff}}=\frac{v_0^2}{2\gamma},$ different from the $n=3$ case.

With this information at hand, we look at the full time-dependent position distribution $P (x,t) $ in terms of  $P_\alpha (x,t) ,$ the probability that the particle is at position $x$ at time $t$ and $\sigma_{x}=\alpha$.
The corresponding Fokker-Planck equations are,
\bea
\frac{\partial P_+}{\partial t} &=& - v_0 \frac{\partial P_+}{\partial x} - \gamma P_+ + \frac \gamma 2 P_0, \cr
\frac{\partial P_-}{\partial t} &=&  v_0 \frac{\partial P_-}{\partial x} - \gamma P_- + \frac \gamma 2 P_0, \cr
\frac{\partial P_0}{\partial t} &=&   - \gamma P_0 + \gamma   (P_+ + P_-) .
\label{EQ:4stateFP}
\eea
We write $P_{\pm 1}$ as $P_{\pm}$ and drop the arguments of $P_{\alpha}$s for brevity.

We choose the initial conditions such that all $\sigma$ values are equally likely, \ie ,
\bea
P_0 (x,0) =\frac{1}{2}\delta (x) ,\cr
P_{\pm} (x,0) =\frac{1}{4}\delta (x).
\label{EQ:4st_ic}
\eea

To solve Eqs.~\eqref{EQ:4stateFP}, it is convenient to introduce the Fourier transform  of $P_{\alpha}(x)$ with respect to $x$, \ie, $\tilde{P}_{\alpha} (k) =\int_{-\infty}^{\infty} e^{ikx}P_{\alpha} (x) d x$. Upon doing the Fourier transform, Eqs.~\eqref{EQ:4stateFP} reduce to a set of coupled ordinary differential equations,
\bea
\frac{\partial}{\partial t}\bar{P}&=&\Omega\bar{P}
\label{4st_matrixfp}
\eea
where, $$\bar{P}=\begin{bmatrix}
\tilde{P}_+ \\ \tilde{P}_- \\ \tilde{P}_0
\end{bmatrix};~~ \text{ }
\Omega=\begin{bmatrix}
-\gamma+i k v_0&& 0&&\gamma/2\\
0 && -\gamma-i k v_0 && \gamma/2\\
\gamma && \gamma && -\gamma
\end{bmatrix}.$$
The solution of the set of equations  Eq.~\eqref{4st_matrixfp} can be written in terms of the eigenvalues and eigenvectors of the matrix $\Omega$,
\bea
\tilde{P} (k,t) &=&e^{-\gamma t}\bigg (a_0\bar{A}_0+a_+e^{\lambda t}\bar{A}_++a_-e^{-\lambda t}\bar{A}_-\bigg) 
\label{4statefourier1}
\eea
where we have used the fact that the eigenvalues of $\Omega$ are $-\gamma,-\gamma\pm \lambda$, with $\lambda=\sqrt{\gamma^2-k^2v_0^2}$.  $\bar A_{0,\pm}$ are the corresponding eigenvectors,
\bea
\bar{A}_0=\begin{bmatrix}
\frac{i\gamma}{2kv_0}\\
-\frac{i\gamma}{2kv_0}\\
1
\end{bmatrix};~~
\bar{A}_{\pm}=\begin{bmatrix}
\frac{ikv_0\pm\sqrt{\gamma^2-k^2 v_0^2}}{2\gamma}\\
\frac{ikv_0\mp\sqrt{\gamma^2-k^2 v_0^2}}{2\gamma}\\
1
\end{bmatrix}.
\eea

The coefficients $a_{\alpha}$s can be determined using the intial conditions Eq.~\eqref{EQ:4st_ic},
\bea
a_0=\frac{-q^2}{2 (1-q^2) };\quad a_{\pm}=\frac{1\pm\sqrt{1-q^2}}{4 (1-q^2) },
\eea
with $q=\frac{k v_0}{\gamma}$. {Substituting these coefficients in Eq.~\eqref{4statefourier1}, we get,
\bea
\tilde P(k=\frac{\gamma q}{v_0},t) =&\frac{e^{-\gamma t}}{2 (q^2-1) }\bigg (-q^2+ (2-q^2) \cosh (\gamma t\sqrt{1-q^2}) \nonumber\\&+2\sqrt{1-q^2}\sinh (\gamma t\sqrt{1-q^2}) \bigg). 
\label{Eq.4statelaplace}
\eea
Eq.~\eqref{Eq.4statelaplace} can be inverted exactly using Bessel Function identities
[The Fourier inversion is carried out in detail in Appendix \ref{4stFourierInv}.
\bea
P (z= \frac{x}{v_0t},t) &=& \gamma t\frac{e^{-\gamma t}}{2}\bigg[\frac{1}{\sqrt{1-z^2}}I_1 (\gamma t\sqrt{1-z^2}) + I_0 (\gamma t\sqrt{1-z^2})\cr
&-&  \frac{\gamma t |z|}{4} - \frac{1}{2\gamma t}\int_{|z|}^{1} d\omega (\frac{\partial^2}{\partial z^2}) I_0 (\gamma t\sqrt{\omega ^2-z^2}) \bigg]\cr
&+& \frac{ e^{-\gamma t}}{4 v_0 t}\bigg (2\delta (z) +\delta (z-1) +\delta (z+1)\bigg).
\label{EQ:4stfulldistribution}
\eea
Note that this solution is valid for $|z| < 1$, $P(z,t)$ is zero otherwise. The integral in the above equation can be evaluated numerically to arbitrary accuracy for any $x.$ $P(x,t)$ obtained from  Eq.~\eqref{EQ:4stfulldistribution} is compared with numerical simulations in  
Fig.~\ref{fig:4st_px}~(a) for $\gamma=1$ and different values of $t$ which show an excellent match.

\begin{figure*}[t]
 \centering
 \includegraphics[width=8 cm]{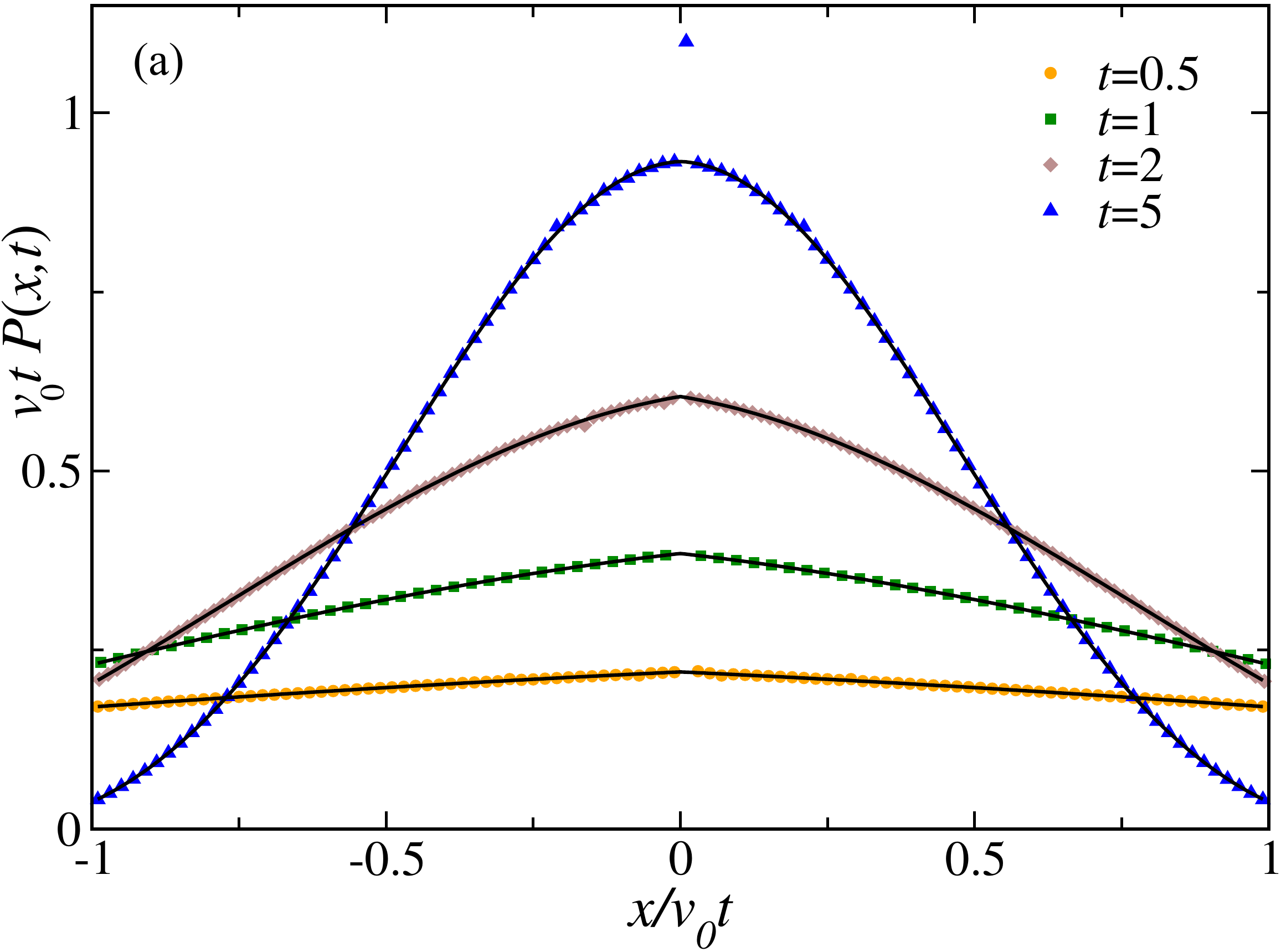}~~~~~
 \includegraphics[width=8 cm]{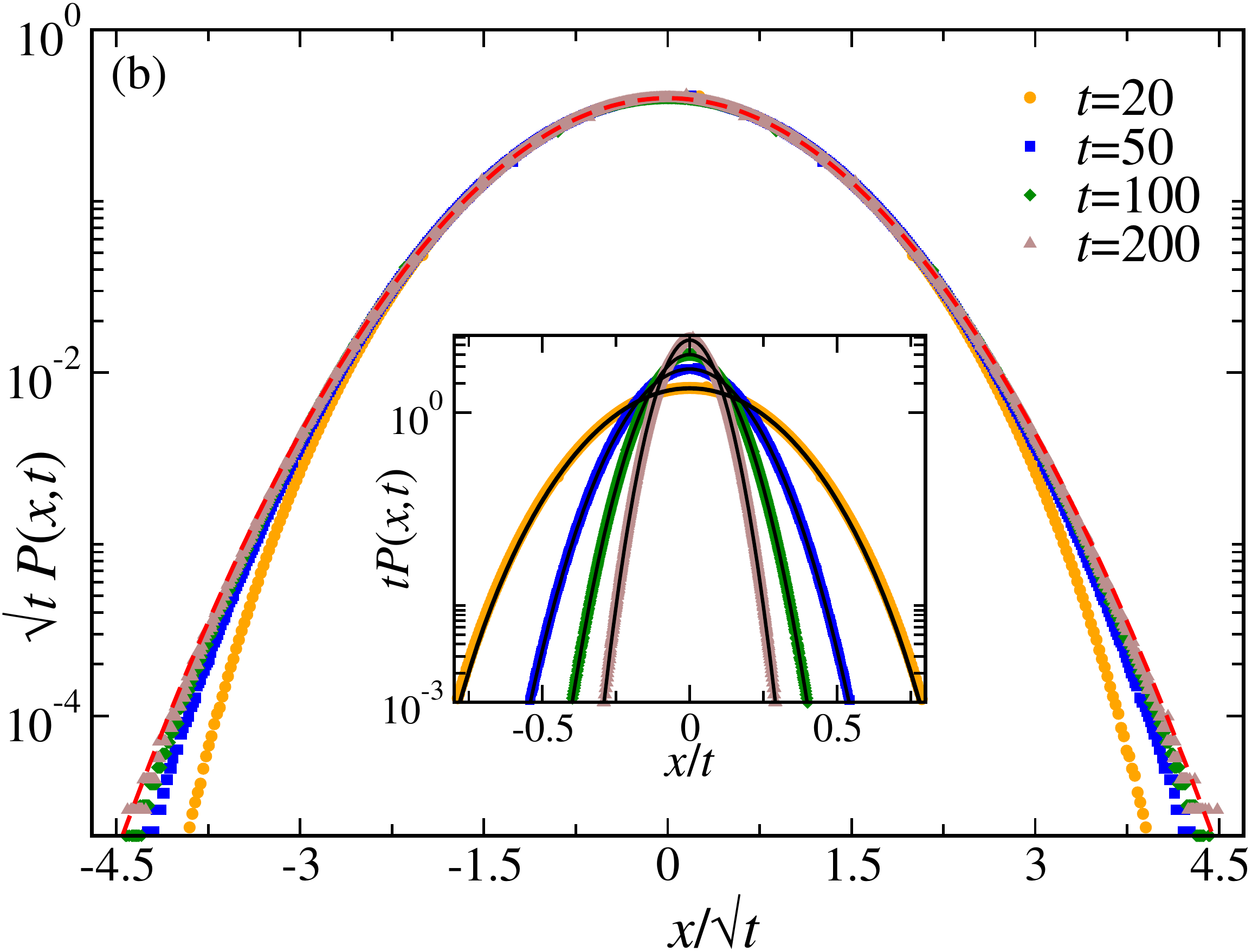}
 \caption{4-state model: (a) Plot of $x$-marginal for different values of t and $\gamma=1.0$  The solid black lines correspond to the analytical prediction Eq.~\eqref{EQ:4stfulldistribution} and the symbols correspond to the data from numerical simulations. For better visibility we have excluded the delta functions at the origin and the boundaries.
(b) Plot of $P(x,t)$ obtained from numerical simulations, as a function of the scaled variable $w=x/\sqrt{t}$ for different (large) values of $t$ and $\gamma=1.$ The red dashed lines shows the corresponding Gaussian distribution. The inset shows the same data as a function of $x/t.$  The solid black lines there correspond to the analytical prediction Eq.~\eqref{4st_x_ld}.  Here we have used $v_0=1.$ 
 }
 \label{fig:4st_px}
\end{figure*}

The asymptotic forms of the distribution are easy to calculate from Eq.~\eqref{EQ:4stfulldistribution}. At short times ($t\ll\gamma^{-1}$), the distribution is dominated by the three $\delta-$functions at $0,\pm v_0t$ while in the bulk it is linear,
\bea
P (z,t) &\approx & \frac{\gamma t}{2} \left(1-\frac{\gamma t|z|}{2}-\frac{\gamma t}{4} \right). 
\eea
At long times ($t\gg\gamma^{-1}$),  using the asymptotic expressions for the modified Bessel functions $I_0$ and $I_1$, we get a large deviation form,
\bea
P (z,t) &=&\frac{v_0  e^{-\gamma t (1-\sqrt{1-z^2}) }}{\sqrt{4\pi D_{\text{eff}}/t}}
\label{4st_x_ld}
\eea
with the large deviation function
\bea
\phi(z)=\gamma(1-\sqrt{1-z^2}).
\eea 
The typical fluctuations in $x$ are $\sim\sqrt{t}$ and Gaussian in nature. Thus the distribution near the origin can be written in terms of the scaled variable $w=x /\sqrt{t}$ as 
\bea
P(w,t)\simeq \frac{1}{\sqrt{4\pi D_{\text{eff}}}}\exp\left(-\frac{\gamma w^2}{2v_0^2 }\right). 
\eea
A comparison of the obtained large deviation form, Eq.~\eqref{4st_x_ld}  (solid lines)  and numerical simulation is shown in Fig.~\ref{fig:4st_px}~(b) inset for $t=100$ and three values of $\gamma$. Figure.~\ref{fig:4st_px}~(b) shows a plot of $P(w,t)$ as a function of the scaled variable $w$, a collapse is seen near the peak at $w=0$. However the tails are non Gaussian and do not collapse.

Thus in this model, we see again the short time distribution  dominated by three $\delta-$functions and linear in the bulk. This $3-$peaked structure evolves in time to a single Gaussian like peak at the centre.

\section{Continuous $\theta$}\label{sec:cont}

In this section we consider the case where the orientation of the RTP is a continuous variable and can take any real values in the range $[0, 2 \pi]$, \ie, the particle travels at a constant speed $v_0$, along the direction $\hat{n}= (\cos\theta ,\sin\theta) $ until it flips and changes its orientation to a new $\theta '$, it then moves with the same constant speed $v_0$ along the new orientation $\theta'$. The rate of this flipping is $\gamma$, while the new orientation is chosen from a uniform distribution $\in [0,2\pi]$. Thus the effective rate of flipping from $\theta\rightarrow\theta'$ is given by $\frac{\gamma}{2\pi}$,see Fig.~\ref{fig:schem_all} (c). A typical trajectory following such dynamics is illustrated in Fig.~\ref{fig:schem_all} (f).

The time evolution of the $2$d distribution obtained from numerical simulations is shown in Figs.~\ref{fig:2dcont}(g), (h) and (i).The distribution is isotropic at all times, however at times  $t \ll \gamma ^{-1}$, the particles crowd away from the origin taking the form of a circle or radius $v_0 t$. This marks the boundary of the distribution in the $x-y$ plane. As time increases the crowding at the boundary decreases and the origin starts populating as is evident from Fig.~\ref{fig:2dcont}(h). Finally we get a centrally peaked distribution at times larger than $\gamma^{-1}$ (Fig.~\ref{fig:2dcont}(i)).

This model has been studied previously \cite{Stadje,Martens2012}, where exact expressions for the radial distribution is obtained. We present a simpler derivation leading to the same results and then go on to discuss the exact and large deviation form of the  marginal distribution which shows some intriguing behaviour.

\subsection{Moments and Cumulants}
 Let us first look at the moments to see the short and long time behaviour of the particle. We assume that the initially the particle is oriented along a random direction $\theta_0 \in [0,2 \pi]$ with probability $\frac 1{2\pi}.$   The new orientation at each tumble event is also chosen from a uniform distribution in $[0,2\pi]$; because of this rotational symmetry the $x$ and $y$ directions are equivalent and the odd moments are zero at all times. The first non-zero moment, the variance can be calculated using the 2-point $\sigma$ correlations (See Appendix \ref{sec:app_theta}).
 \bea
 \langle x^2(t)\rangle &=&\frac{v_0^2}{\gamma}\left ( t-\frac{1-e^{-\gamma t}}{\gamma}\right)
 \label{mcont}
 \eea
 Thus, at short times ($t\ll\gamma^{-1}$) , 
 \bea
 \langle  x^2(t) \rangle = v_0^2 t^2+O(t^3) \n
 \eea
 which indicates that the motion is ballistic in this regime. This goes over to being diffusive at large times ($t\gg\gamma^{-1}$) ,
 \bea
 \langle  x^2(t) \rangle \simeq 2D_{\text{eff}}t
 \eea
  with $D_\text{eff}=\frac{v_0^2}{2\gamma}$. Thus we see that the behaviour of this model is qualitatively same as the two discrete models considered in the previous sections.
  
\subsection{Position Distribution}

Let us consider that the particle begins from origin at $t=0$, pointing along $\hat{n}_0= (\cos\theta_0,\sin\theta_0) $, where $\theta_0$ can be any angle between $[0,2\pi]$, then $\cal{P} (\vec{r},\theta,t|\theta_0) $ denotes the probability for the particle to be at $ (\vec{r},\theta) $ at time $t$, given $\theta_0$. It evolves according to the Fokker-Planck equation,
 \bea
 \frac{\partial}{\partial t} \cal{P} (\vec{r},\theta,t|\theta _0) &=&-v_0\hat{n}.\vec{\nabla}\cal{P} (\vec{r},\theta,t |\theta_0) -\gamma \cal{P} (\vec{r},\theta,t|\theta_0) \nonumber\\&+&\gamma\int\frac{d\theta'}{2\pi}\cal{P} (\vec{r},\theta',t|\theta_0) ,
 \label{fpcontinuous}    
 \eea 
 where the first term on the right is the drift term, the second term is the probability that the RTP can flip to some other orientation at rate $\gamma$, while the third term takes into account that the RTP can flip to $\theta$ from any other $\theta '$ in $[0,2\pi]$. Let us define the Fourier-Laplace transform of $\cal{P} (\vec{r},\theta,t|\theta _0) $,
 \bea
 \hat{\cal{P}}(\vec{k},\theta,s\mid\theta_0) =\int_0^{\infty}dt~e^{- s t}\int d\vec{r}~e^{i\vec{k}.\vec{r}} \cal{P} (\vec{r},\theta,s\mid\theta_0)\;
\label{defineFTLTcont} 
 \eea
 where $\vec{k}= (k_1,k_2) $.  We need to solve Eq.~\eqref{fpcontinuous} with the initial condition,
 \bea
 \cal{P} (\vec{r},\theta,0|\theta_0) =\delta ^2 (\vec{r})\delta (\theta-\theta_0) 
 \label{cont_ic} 
 \eea
where $\theta_0$ is some arbitrary angle in $[0,2\pi]$.  Using Eq.~\eqref{cont_ic} and Eq.~\eqref{defineFTLTcont}, Eq.~\eqref{fpcontinuous} becomes, 
 \bea
 s\hat{\cal{P}} (\vec{k},\theta,s\mid\theta_0) =\delta  (\theta -\theta_0) + i v_0\vec{k}.\hat{n} \hat{\cal{P}} (\vec{k},\theta,s\mid\theta_0) \nonumber \\ -\gamma \hat{\cal{P}} (\vec{k},\theta,s\mid\theta_0) +\gamma \int \frac{d\theta'}{2\pi}\hat{\cal{P}} (\vec{k},\theta',s\mid\theta_0).
 \eea
Solving for $\hat{\cal{P}} (\vec{k},\theta,s\mid\theta_0) $, we have,
\bea
\hat{\cal{P}} (\vec{k},\theta,s\mid\theta_0) &=&\frac{\bigg (\delta (\theta -\theta_0) +\gamma\int_0^{2\pi}\frac{d\theta'}{2\pi}\hat{\cal{P}} (\vec{k},\theta',s\mid\theta_0) \bigg)  }{s+\gamma-i v_0\vec{k}.\hat{n}}. \n
\label{Pkthetastheta0}
\eea

Integrating over the final and initial orientations $\theta$ and $\theta_0$, Eq.~\eqref{Pkthetastheta0} reduces to an algebraic equation,
 \bea
 G (\vec{k},s) &=&f (\vec{k},s) +\gamma G (\vec{k},s) f (\vec{k},s), 
 \label{contgfrelation}
 \eea
where
\bea
G (\vec{k},s) &=&\int_0^{2\pi}d\theta\int_0^{2\pi}\frac{d\theta_0}{2\pi}\hat{\cal{P}} (\vec{k},\theta,s\mid\theta_0),
\label{gksdefn}
\eea 
and, 
\bea
 f (\vec{k},s) &=&\int_0^{2\pi}\frac{d\theta}{2\pi}\frac{1}{s+\gamma-i v_0\vec{k}.\hat{n}}=\frac{1}{\sqrt{ (\gamma +s) ^2+v_0^2 k ^2}},\;\;\label{EQ:fks}
\eea 
with $k^2=k_1^2+k_2^2.$ From Eq.~\eqref{contgfrelation} and  Eq.~\eqref{EQ:fks} we get,
\bea
G(\vec{k},s) &=&\frac{f (\vec{k},s) }{1-\gamma f (\vec{k},s) }=\frac{1}{\sqrt{ (\gamma +s) ^2+v_0^2 k ^2}-\gamma}. \label{EQ:gfcontinuouslambdaks}
 \eea

Before proceeding further, let us first note that, $G (\vec k,s),$ by its definition Eq.~\eqref{gksdefn}, is the Fourier-Laplace transform of the full distribution function $\cal{P} (x,y,t) $, so any moment of the position in the $s-$space can be obtained by taking derivatives of Eq.~\eqref{EQ:gfcontinuouslambdaks} with respect to either $ik_1$ or $ik_2$ at $k_1=0,k_2=0.$ For example,   
 \bea
 \int_{0}^{\infty} e^{-st}\langle x^2\rangle~d t &=&\left. \frac{\partial^2 G (\vec k,s) }{\partial (ik_1)^2} \right |_{k_1=0 \atop k_2=0}=\frac{v_0^2}{s^2  (\gamma+s) }.\nonumber
 \eea
 This can be inverted to compute the second moment of the $x$-component, $\la x^2(t)\ra,$ and matches exactly with Eq.~\eqref{mcont}.
 
Now, to obtain the position distribution, we need to find the Laplace-Fourier inverse of $G(\vec k,s).$ For simplicity, we drop the vector notation of $k$ in $G$ and $f$ henceforth, as both depend only on $k^2$. $G (k,s) $ has contributions from all the events where the particle does not flip or flips multiple times till time $t.$ It turns out that to invert Eq.~\eqref{EQ:gfcontinuouslambdaks} it is convenient if we subtract the contribution of the no flip event, from $G (k,s) $. This contribution can be calculated explicitly (See Appendix \ref{nojump})  and comes out to be equal to $f (k,s) $. We define,
\bea
\cal{G} (k,s) &=&G (k,s) -f (k,s). 
\label{calgks}
\eea
 The inversion of $\cal{G}(k,s)$ is non-trivial and has been carried out in details in Appendix \ref{fourlapinver}. The resulting contribution to the probability distribution is
\begin{figure*}[t]
 \centering
 \includegraphics[width=8 cm]{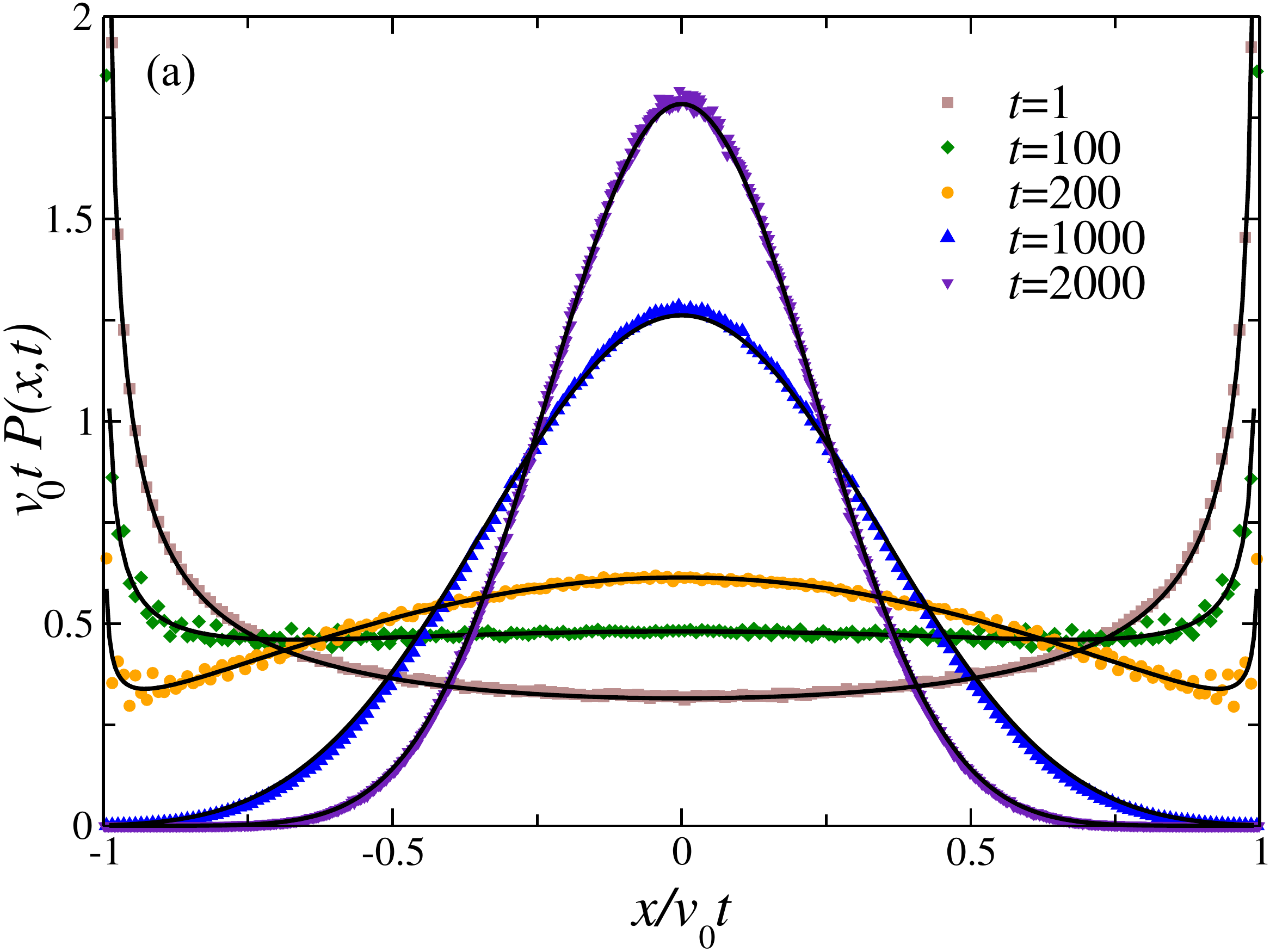}~~~~~
 \includegraphics[width=8 cm]{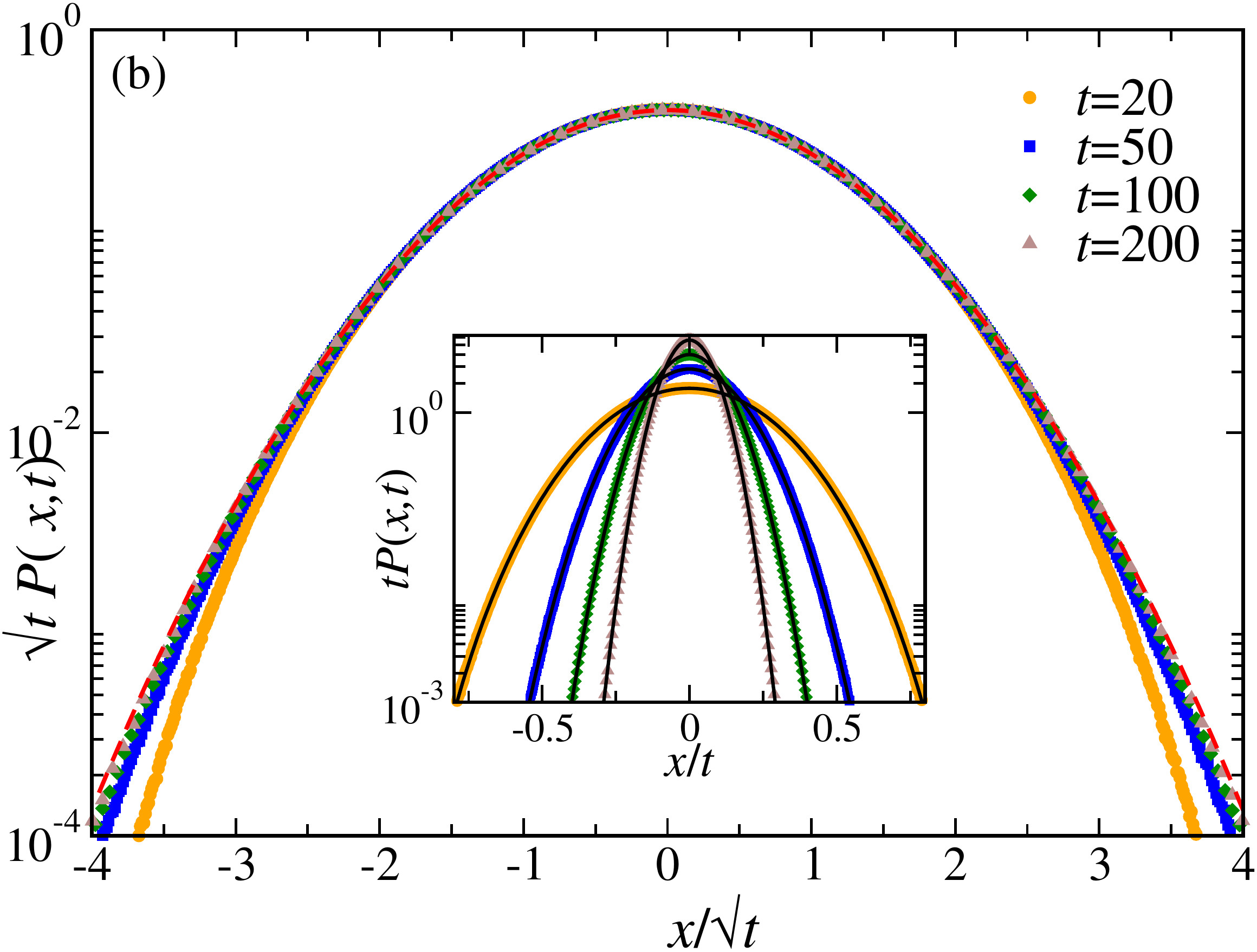}
 \caption{Continuous $\theta$ model: (a) Plot of $x$-marginal for different values of t and $\gamma=1.0$  The solid black lines correspond to the analytical prediction Eq.~\eqref{cont_full} and the symbols correspond to the data from numerical simulations. For better visibility we have excluded the delta functions at the origin and the boundaries.
(b) Plot of $P(x,t)$ obtained from numerical simulations, as a function of the scaled variable $w=x/\sqrt{t}$ for different (large) values of $t$ and $\gamma=1.$ The red dashed line shows the corresponding Gaussian distribution. The inset shows the same data as a function of $x/t.$  The solid black lines there correspond to the analytical prediction Eq.~\eqref{contld}.  Here we have used $v_0=1.$ 
 }
 \label{fig:cont}
\end{figure*}

\bea
 P (r,t) &=&\frac{\gamma e^{-\gamma t}}{2\pi v_0}\frac{\exp\bigg[\frac{\gamma}{v_0} \sqrt{v_0^2 t^2-r^2}\bigg]}{\sqrt{v_0^2 t^2-r^2}}.
 \eea
To get the full distribution we have to add the contribution of the no-flip event to the above equation. That contribution is calculated in Appendix \ref{nojump}
\bea
P_0 (r,t) &=&\frac{e^{-\gamma t}}{2\pi r}\delta (r-v_0 t). \label{eq:_P0rt}
\eea 
Thus we have the exact position distribution at any time $t,$
 \bea 
 \cal{P} (r,t) &=&e^{-\gamma t}\bigg[\frac{\delta (r-v_0 t) }{2\pi r}\cr
 &+&\frac{\gamma }{2\pi v_0}\frac{\exp\bigg[\frac{\gamma}{v_0} \sqrt{v_0^2 t^2-r^2}\bigg]}{\sqrt{v_0^2 t^2-r^2}}\Theta (v_0 t-r) \bigg].~
 \label{EQ:continuousfullr}
 \eea
The $\Theta$ function implies that the distribution is always bounded. This expression is identical to the ones obtained in \cite{Stadje, Martens2012}.

 \subsection{Marginal Distribution}

We now look at the marginal distribution along either $x$ or $y$. For this purpose, we rewrite Eq.~\eqref{EQ:continuousfullr} in terms of the Cartesian coordinates as,
\bea
\cal{P} (x,y,t) &=&e^{-\gamma t}\bigg[\frac{\delta (\sqrt{x^2+y^2}-v_0 t) }{2\pi \sqrt{x^2+y^2}}\cr &+&\frac{\gamma e^{\left (\frac{\gamma}{v_0} \sqrt{v_0^2 t^2-x^2-y^2}\right) }}{2\pi v_0 \sqrt{v_0^2 t^2-x^2-y^2}}\Theta (v_0 t-\sqrt{x^2+y^2}) \bigg].~~~
\eea
The marginal distribution of $x$ is obtained by integrating over $y,$ \ie,  $ P (x,t)=\int_{-\infty}^{\infty}dy~\cal{P} (x,y,t)$, which yields
 \bea
 P (x,t) &=&\frac{\gamma e^{-\gamma t}}{2v_0} \Bigg( L_0 \left[\frac{\gamma}{v_0} W(x)\right] +I_0 \left[\frac{\gamma}{v_0} W\right]  \Bigg)+\frac{e^{-\gamma t}}{\pi W(x)}~~
\label{cont_full}
 \eea
 where $W(x)=\sqrt{v_0^2t^2-x^2};$ $ {I_0}$ is the the modified Bessel function of the first kind  and $ {L_0}$ is the modified Struve function~\cite{dlmf}.
 
 The interesting difference between the marginal distribution of this model and the two previously discussed discrete models is that the divergence at the boundaries is not a $\delta-$function divergence but an algebraic divergence. The analytic expression of the distribution function found in Eq.~\eqref{cont_full} is compared with numerical simulation for  $\gamma=0.01$ for different values of $t$ in Fig.~\ref{fig:cont}(a).

 We can immediately look at the asymptotic limits of the distribution, using the asymptotic forms of the modified Bessel and modified Struve functions \cite{dlmf}, where the active and passive characteristics are more prominent.  At very short times  ($t\ll\gamma^{-1}$), the distribution  is dominated by the no flip process, given by
\bea
P (z=\frac{x}{v_0 t},t) & \approx &\frac{e^{-\gamma t}}{\pi \sqrt{1-z^2}},
\eea 
 while at large times ($t\gg\gamma^{-1}$)  we get a large deviation form from the asymptotic expansions of  $ {I_0}$ and $ {L_0}$. Thus,
 \bea
 P (z,t) &\approx&
 \frac{v_0 t~ e^{-\gamma t (1-\sqrt{1-z^2}) }}{\sqrt{4\pi D_{\text{eff}}t}}.
\label{contld} 
 \eea
 with the large deviation function
 \bea
 \phi(z)=\gamma (1-\sqrt{1-z^2}).
 \eea
 We can actually get the above large deviation form directly from Eq.~\eqref{EQ:gfcontinuouslambdaks} by taking a large time approximation to do the inverse time laplace transform and doing a saddle-point approximation thereafter. This calculation is added in Appendix \ref{nojump}.
 
 The large deviation form of the distribution obtained in Eq.~\eqref{contld} is compared with the results of numerical simulation at $t=2000$ for four different values of $\gamma$ in the inset of Fig.~\ref{fig:cont}~(b) 
 The typical fluctuations in $x$ are Gaussian and scale as $\sqrt{t}$. Thus the distribution near the origin in terms of the scaled variable $w=x/\sqrt{t}$ becomes
 \bea
 P(w,t)&\simeq& \frac{1}{\sqrt{4\pi D_{\text{eff}}}}\exp[-\frac{\gamma w^2}{2v_0^2 }]. 
 \eea
Figure~\ref{fig:cont}~(b) shows a plot of $P(w,t)$ with the scaled variable $w$. We see a scaling collapse near the peak while near the boundaries there is no collapse indicating non-Gaussian tails.
 
 Summarizing, we see at short times, this model is dominated by the divergence at the boundaries, like the discrete models described in the previous two sections. However here, the nature of divergence is algebraic unlike the $\delta-$functions of the previous two models. This short time $2-$peaked distribution goes over to a single Gaussian like peak at large times.

\section{Conclusion} \label{sec:conclusions}

We have studied a set of RTP models in two spatial dimensions, 
where the orientation $\theta$ of the particle can take either discrete or continuous values. We show that, in all the cases, the flipping rate of the orientation provides a time-scale which separates two very different dynamical regimes. In the short-time regime, 
the RTPs show an `active' ballistic behaviour, with a model-specific effective velocity. This  active regime is also characterized by non-trivial position distributions, which we compute exactly. 
It turns out that, the shape of the position distributions in this short-time regime also depends crucially on the specific dynamics. 
On the other hand, in the long-time regime, all the models show an effective diffusion-like behaviour where the typical position fluctuations are characterized by Gaussian distributions, albeit the width of the distribution depends on the microscopic dynamics. 
However, we show that the signature of the activity retains itself in the atypical fluctuations. We compute the large deviation functions explicitly which, as expected, also depends on the specific model. 

 Previously RTPs have been studied in higher dimensions, but very few analytical results were available in literature, mostly focusing on the diffusivity. The RTP dynamics considered in this article are simple models which lend themselves easily to analytical treatments starting from the microscopic dynamics. The analytical results obtained here point to a very generic qualitative behaviour of the RTPs, a short-time ballistic regime where we see a crowding at the boundaries while a long-time diffusive regime where the gathering is around the origin.
 We believe our work will be informative for the study of other active particle dynamics in higher dimensions with more complexities like added rotational diffusion. Possible extensions can be to ask other typical questions related to active motion, \eg, first passage properties and behaviour in the presence of external confinements, in the context of these models. It would be also interesting to verify some of our analytical predictions in experimental systems.

\begin{acknowledgements}

U. B. acknowledges support from Science and Engineering Research Board  (SERB) , India under Ramanujan Fellowship  (Grant No.  SB/S2/RJN-077/2018). 
 
\end{acknowledgements}

\appendix

\section{Calculation of the Propagator for the $\theta$ Processes and $2$-point $\sigma$ Correlations} \label{sec:app_theta}

\subsection*{$n$-state Model}

Let $\theta_j=\frac{2\pi j}{n}, j=0,1, \dots n-1$ denote the $n$ possible values of $\theta$, and $P_j(t)$ denote the probability that the particle orientation is $\theta_j$ at time $t$. The  Fokker Planck equation governing the time-evolution of $P_j(t)$ with periodic boundary conditions, $P_n(t)=P_0(t)$, is
\bea
\frac{d } {d t} P_j &=&-\gamma P_j+\frac{\gamma}{2}P_{j+1}+\frac{\gamma}{2}P_{j-1}.
\eea
 This set of equations is easily solved by  going to the Fourier basis,
\bea
P_j(t)&=&\sum_{k=0}^{n-1}e^{i\frac{2\pi jk}{n}}Q_k(t)
\eea
where $Q_k(t)=\frac 1 n\sum_{j=0}^{n-1}e^{-i\frac{2\pi jk}{n}}P_j(t)$.The time dependence of $Q_k$ is given by
\bea
Q_k(t)=Q_k(0)~e^{-\lambda_k t}
\eea
where, $\lambda_k=\gamma(1-\cos \frac{2\pi k}{n})$ are the eigenvalues of the tri-diagonal matrix.
Now, with initial conditions, $P_j(0)=\delta_{jm}$; $\theta(0)=\frac{2\pi m}{n}$, we have
\bea
P_j(t)&=&\frac 1 n\sum_{k=0}^{n-1}~e^{i\frac{2\pi k(j-m)}{n}}~e^{-\lambda_k t}.
\eea
Thus we can write the propagator of the process starting with $\theta_0$ at time $t=0$ as
\bea
P(\theta,t|\theta_0,0)&=&\frac 1 n\sum_{k=0}^{n-1}e^{ik(\theta-\theta_0)} e^{-\lambda_k t}.
\eea 
Using this we can calculate the $2$ or higher point correlations of the $\sigma$s defined in main text.For example,
\bea
\langle \sigma_x(t) \sigma_x(0)\rangle &=&\frac{1}{n}\sum_{\theta,\theta_0}\cos \theta \cos\theta_0~ P(\theta,t|\theta_0,0)\nonumber \\
&=&\frac{1}{n^2}\sum_{k=0}^{n-1} e^{-\lambda_k t}|\sum_{\theta}\cos\theta e^{ik\theta}|^2.
\eea

\subsection*{Continuous Model}
Here the propagator can be written as a sum of contributions from events where the final and initial $\theta$ are same (\ie, no $\theta$ flip) and where they are different. They can be as in Eq.~\eqref{cont1}
\bea
P(\theta,t|\theta_0,0)&=&e^{-\gamma t}\delta(\theta-\theta_0)+(1-e^{-\gamma  t})\frac{1}{2\pi}.
\eea
Thus the $2-$point $\sigma$ correlations can be evaluated as
\bea
\langle \sigma_i(t)\sigma_i(0)\rangle&=&\frac{1}{2\pi}\int d\theta ~d\theta_0 P(\theta,t|\theta_0,0)\sigma_i(t)\sigma_i(0).~
\eea
Now, $\sigma_\alpha(t)$ is $\cos \theta$ or $\sin \theta$ for $\alpha=x$ and $\alpha=y$ respectively. Using the properties of $\sin$ and $\cos$ functions, the integral contributes only when there is no $\theta$ flip. Thus we have,
\bea
\langle \sigma_x(t)\sigma_x(0)\rangle &=&e^{-\gamma t}\frac{1}{2\pi}\int_0^{2\pi}d\theta~\cos ^2\theta =\frac{1}{2}e^{-\gamma t}\nonumber\\
\langle \sigma_y(t)\sigma_y(0)\rangle &=&e^{-\gamma t}\frac{1}{2\pi}\int_0^{2\pi}d\theta~\sin ^2\theta = \frac{1}{2}e^{-\gamma t}.\nonumber
\eea

\section{Details of the 3-state X Marginal Distribution}\label{sec:app_int}

 We rewrite Eqs.~\eqref{EQ:3stxlaplace} in the main text for  $x\ne 0$,
 \bea
v_0\frac{d P}{d x}&=& W P.
\label{appmatrix}
 \eea
where, 
\bea
P=\begin{bmatrix}
\hat P_+\\
\hat P_-
\end{bmatrix}\text{ and }W=\begin{bmatrix}
-(s+\gamma) && \gamma/2\\
-2\gamma && (2s+\gamma)
\end{bmatrix}.
 \eea
The eigenvalues of $W$ are given by $(s\pm \lambda)/2$, where $\lambda=\sqrt{3s (3s+4\gamma) }$. Using these eigenvalues and implementing the  boundary conditions that $\hat P_{\pm}(x,s)\rightarrow 0$ as  $x\rightarrow\pm\infty $, we have
 \bea
 \hat P_+ (x,s)   = \left \{ \begin{split}
                         A_+  \exp \left[-\frac x{2v_0} (\lambda-s) \right] & \quad \text{for}~ x>0 \cr
                         B_+ \exp \left[\frac x{2v_0} (\lambda+s) \right]~ & \quad \text{for}~ x<0
                         \end{split}
\right. \label{eq:3st_Php}
\eea 
and
\bea
\hat P_- (x,s)   = \left \{ \begin{split}
                         A_- \exp \left[-\frac x{2v_0} (\lambda-s) \right] & \text{for}~ x>0 \cr
                         B_- \exp \left[\frac x{2v_0} (\lambda+s) \right]~ &  \text{for}~ x<0,
                         \end{split}
\right.   \label{eq:3st_Phm}                      
\eea
 where and $A_{\pm}$ and $B_{\pm}$ are arbitrary constants. Putting $P_{\pm}$ back in Eq.~\eqref{appmatrix}, we have,
 \bea
 A_-&=&\frac{A_+ (2\gamma + 3 s -\lambda) }{\gamma},\nonumber\\
 B_-&=&\frac{B_+ (2\gamma + 3 s +\lambda)  }{\gamma}. 
 \eea 
Next, to evaluate the constants $A_+$ and $B_+$, we note that due to the presence of the $\delta$-functions, integrating  the original Eqs.~\eqref{EQ:3stxlaplace}  around the origin $x=0$ yields  discontinuity conditions for $\hat{P}_\pm$ across $x=0$, 
\begin{align*}
&v_0 \left[A_+ -B_+ \right] =\frac 13, \\
&v_0 \left[\frac{A_+ (2\gamma + 3 s -\lambda) }{\gamma}  -  \frac{B_+ (2\gamma + 3 s +\lambda)  }{\gamma} \right] =-\frac 43.
\end{align*}
Solving these two equations determines the constants as
 \bea
 A_+= \frac{3 (2\gamma+s) +\lambda}{6 v_0 \lambda},~ B_+= \frac{3 (2\gamma+s) -\lambda}{6 v_0 \lambda}. \label{eq:3st_AB}
 \eea
 Using Eq.~\eqref{eq:3st_AB} in Eqs.~\eqref{eq:3st_Php} and \eqref{eq:3st_Phm} and adding $\hat P_+$ and $\hat P_-,$ we get 
the Laplace transform of the position distribution $P (x,t) $ as given by Eq.~\eqref{EQ:3statexlaplace} in the main text.  

Next we show the computation of the inverse laplace transform of Eq.~\eqref{EQ:3statexlaplace} in the main text. Let us consider the case $x>0$. We need to compute the Bromwich integral,
\begin{align}
P (x,t) &=\frac{1}{2\pi i}\int_{c_0-i\infty}^{c_0+i\infty}e^{st}\frac{6\gamma+5s-\lambda}{2v_0\lambda}\exp\left[-\frac{x}{2v_0} (\lambda-s) \right]
ds
\label{bromwich}
\end{align}
where $\lambda=3\sqrt{s (s+a)}$ with $a=4\gamma/3$. The integrand has a branch-cut along the real axis from $s=0$ to $s=-a$, so we draw a contour  keeping the branch-cut to the left of $c_0$, as shown in Fig.~\ref{fig:contour}. This contour can be broken into $6$ different parts as indicated in the figure.

\begin{figure}[ht]
 \includegraphics[width=6 cm]{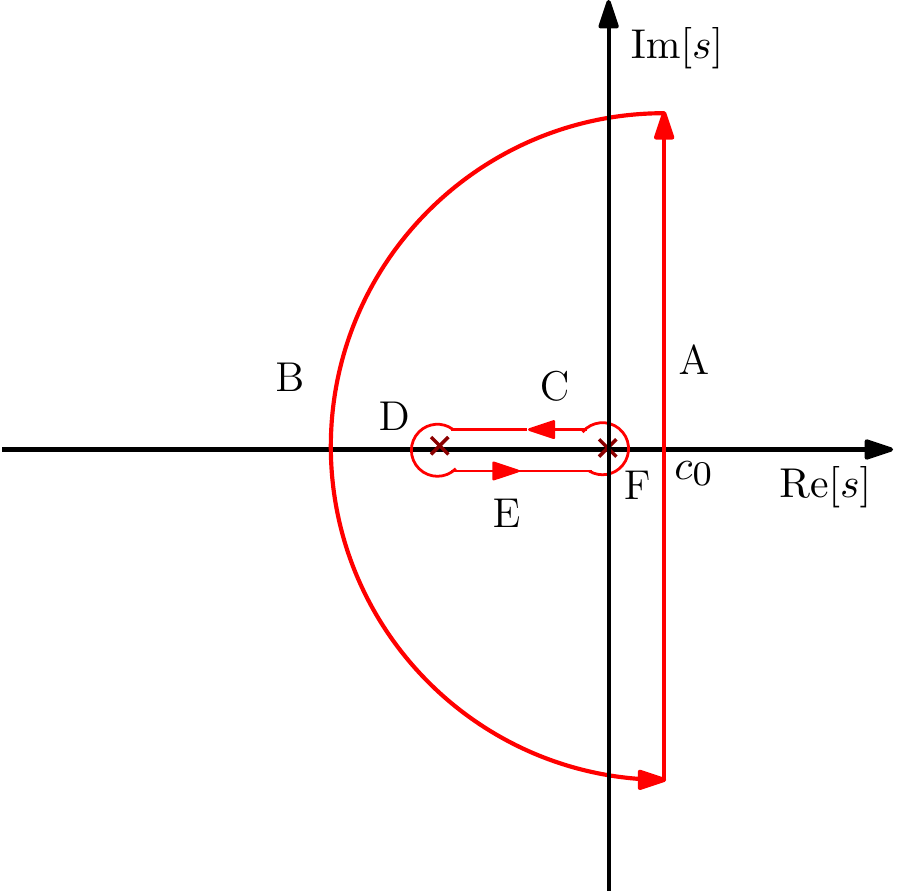}
 \caption{Illustration of the contour used to evaluate the integral \eqref{bromwich}.}
 \label{fig:contour}
\end{figure}
Using Cauchy's theorem, for this contour integral we can write 
\begin{align}
\int_{A}+\int_{B}=\int_{C}+\int_{D}+\int_{E}+\int_{F}. \label{eq:contour_sum}
\end{align}
Now, $\int_{A}=P (x,t) $ is the integral that we need, with $A$ extending to $\infty$.

We first compute the contributions coming from the small circular arcs $D$ and $F$ of radius $\epsilon(\rightarrow 0).$   Along $D$, $s=-a+\epsilon e^{i\theta}$, while along $F$,  $s=\epsilon e^{i\theta}$. It can be immediately seen that the contributions of the integrals along these two circular arcs vanish in the limit $\epsilon \to 0.$ In the following we evaluate the integrals along $ C, E$ and $B$ separately.

Along C, $s=u e^{i\pi}+i \epsilon$, hence, $ds=-du$. With $\epsilon\rightarrow 0$, $\lambda=3i\sqrt{u(a-u)}$ here and we have,
\begin{align}
\int_C= \int_{0}^{a}\frac{-du}{2\pi i}e^{-u (t+\frac{x}{2v_0}) }e^{i\frac{3x}{2v_0}\sqrt{u (a-u) }}\bigg (\frac{6\gamma-5u}{i6v_0\sqrt{u (a-u) }}-1\bigg). \n
\end{align}
On the other hand, along E, $s=u e^{-i\pi}-i \epsilon$, and $ds=-du$. In this case, $\lambda=-3i\sqrt{u(a-u)}$ for $\epsilon\rightarrow 0$,  and we have
\begin{align}
\int_E=\int_{a}^{0}\frac{-du}{2\pi i}e^{-u (t+\frac{x}{2v_0}) }e^{-i\frac{3x}{2v_0}\sqrt{u (a-u) }}\bigg (\frac{i(6\gamma-5u)}{6v_0\sqrt{u (a-u) }}-1\bigg). \n
\end{align}
Adding the contributions from the segments $C$ and $E$, we get,
\bea
\int_C+\int_E &= &\int_0^{a} d u\frac{e^{-u (t+\frac{x}{2v_0}) }}{6 \pi v_0} \left[\frac{ (6\gamma - 5u) }{\sqrt{u (a-u) }}\cos \frac{3x}{2 v_0}\sqrt{u (a-u) }\right.\cr
&&\left. - 3\sin \frac{3x}{2 v_0}\sqrt{u (a-u) }\right].\label{eq:int_CE}
\eea

To evaluate the integral along B, we note that, here the real part of $s$ is negative and $|s| \to \infty;$ hence the integral reduces to,
\begin{align}
\int_B\approx \frac{1}{3v_0} \int\frac{ds}{2\pi i}~\exp \left[s (t-\frac{x}{v_0}) \right] e^{-\frac{\gamma x}{v_0}}
\label{showtheta}
\end{align}
where we have used the fact that along $B,$ $|s| \gg\gamma$ to approximate $\left(\frac{6\gamma+5s-\lambda}{2v_0 \lambda}\right)$ as $\frac{1}{3v_0}$ and $\exp\left(-\frac{x}{2v_0}(\lambda-s)\right)$ as $\exp\left(-\frac{x}{v_0}(s+\gamma)\right)$.

Now, since the integrand in Eq.~\eqref{showtheta} does not have any singularity, we can deform the contour B to be along the imaginary axis, and write,

\bea
\int_B&=&\frac{1}{3v_0} \int_{i\infty}^{-i\infty} \frac{ds}{2\pi i}~\exp \left[s (t-\frac{x}{v_0}) \right]e^{-\frac{\gamma x}{v_0}}\cr
&=&-\frac{1}{3}\delta (x-v_0 t)e^{-\frac{\gamma x}{v_0}} \cr
&=&-\frac{e^{-\gamma t}}{3}\delta (x-v_0 t).\label{eq:intB}
\eea
The required integral $P(x,t)= \int_A$ is now obtained using Eq.~\eqref{eq:contour_sum} along with Eqs.~\eqref{eq:int_CE} and \eqref{eq:intB}. The Bromwich integral for $x<0$  can be also be computed following the same procedure. The final expression for $P(x,t)$ is quoted in Eqs.~\eqref{eq:3st_Pxt} and \eqref{eq:3st_Gxt} in the main text.

 For the case of $P(y,t)$ we also proceed similarly. In that case, however,
the contribution of the integrals around the small circles ($D$ and $F$ in the above contour) are  $\sim \epsilon ^{-1/2}$, and thus cannot be ignored.

To carry out the line integral in Eq.~\eqref{eq:3st_Gxt} in the main text, we first make a change of variable $u=a (w+1)/2$, yielding,
\begin{align}
G_x(x=zv_0 t,& t)  = \frac{\gamma}{9 \pi v_0} e^{-\frac{\gamma t}{3} (z+2)}\int_{-1}^{1} d w~e^{- \frac{\gamma t}{3}(z+2)\, w} \notag\\
\times \Bigl[&- 3\sin z \gamma t\sqrt{1-w^2 } + \frac{ (4 - 5 w) }{ \sqrt{1-w^2 }}\cos z \gamma t\sqrt{1-w^2 } \Bigr].
\label{eq:3st_Gxt2}
\end{align}

Next, we use the following identity from Section 4.124, Eq. 1 of \cite{Gradshteyn}
\begin{align}
F(p,q)\equiv \frac{1}{\pi}\int_{-1}^1 d w\, e^{-p w}\frac{\cos (q \sqrt{1-w^2})}{\sqrt{1-w^2}}= I_0(\sqrt{p^2-q^2}).
\end{align}
Using this identity, we further have,
\begin{align*}
&-\frac{1}{\pi}\int_{-1}^1 d w\, e^{-p w}\sin (q \sqrt{1-w^2}) =\frac{\partial F}{\partial q}=-\frac{q\, I_1\left(\sqrt{p^2-q^2}\right)}{\sqrt{p^2-q^2}}\\
&-\frac{1}{\pi}\int_{-1}^1 d w\, w\, e^{-p w}\frac{\cos (q \sqrt{1-w^2})}{\sqrt{1-w^2}}= \frac{\partial F}{\partial p} = \frac{p\, I_1\left(\sqrt{p^2-q^2}\right)}{\sqrt{p^2-q^2}}.
\end{align*}
Using these, Eq.~\eqref{eq:3st_Gxt2} can be evaluated exactly as
\begin{align}
&G_x(x=zv_0t, t)= \frac{\gamma e^{-\frac{\gamma t}{3} (z+2)}}{9 v_0} \Biggl[4 I_0 \left(\frac{2\, \gamma t }{3} \sqrt{(2z+1) (1-z)}\right) \notag\\
&+ \frac{5-2z}{\sqrt{(2z+1) (1-z)}}I_1 \left(\frac{2\, \gamma t }{3} \sqrt{(2z+1) (1-z)}\right) 
\Biggr],
\end{align}
 which is quoted as Eq.~\eqref{eq:Gx-exact} in the main text.
 
\section{Computation of the Inverse Fourier Transform for $4$-state Marginal Distribution}
\label{4stFourierInv}
In this Section we compute the inverse Fourier transform of 
$\tilde P (q,t) $ given by Eq.~\eqref{Eq.4statelaplace} in the main text. To this end, it is first convenient to rewrite it as,
\bea
\tilde{P} (q,t) &=&e^{-\gamma t}\bigg[\frac{q^2}{2 (q^2-1) }\left (\cosh (\gamma t\sqrt{1-q^2}) -1\right) \cr
&+&\cosh (\gamma t\sqrt{1-q^2}) +\frac{\sinh (\gamma t\sqrt{1-q^2}) }{\sqrt{1-q^2}}\bigg].
\eea 
Let us denote the three terms inside the square brackets in the above equation by $T_1,T_2,T_3$.  Note that the Fourier transform of all the three terms are related by taking derivatives or integrals of each other with respect to the arguments ($q$ and $x$).We exploit this and invert $T_1,T_2,T_3$ term by term, separately. We start by evaluating the Fourier inverse of $T_2$, for this we use an integral Bessel function identity from Section 6.645 Eq. 3 of \cite{Gradshteyn},
\bea
 \int_{-1}^1\frac{1}{\sqrt{1-x^2}}e^{-a}I_1 (b\sqrt{1-x^2}) dx &=& \frac{2}{b} (\cosh\sqrt{a^2+b^2}-\cosh a). \nonumber
\eea
 Let $a=i\gamma q t$ and $b=\gamma t$. Then,
 \bea
  \int_{-1}^1\frac{e^{-i q \gamma t}}{\sqrt{1-x^2}}I_1 (\gamma t\sqrt{1-x^2}) dx &=&\frac{2}{\gamma t} [\cosh  (\gamma t\sqrt{1-q^2}) -\cos (\gamma q t)]. \n
 \eea
 We use the scaling, $\gamma t\rightarrow t'$ and $x\rightarrow x' t $,
 \bea
 \cosh (t'\sqrt{1-q^2}) &=&\frac{t'}{2}\int_{-t'}^{t'}dx' e^{-iqx'}\frac{I_1 (\sqrt{t'^2-x'^2}) }{t'^2-x'^2}+\cos (qt). \nonumber 
 \eea
We can actually call $t'$ and $x'$ as $t$ and $x$ without any ambiguity, throughout the calculations and put back the scaling forms at the end.
 Thus,
 \begin{align}
 \mathcal{F}^{-1}[\cosh (t\sqrt{1-q^2}) &=\frac{t I_1 (\sqrt{t^2-x^2}) }{2\sqrt{t^2-x^2}}\Theta (t-|x|) \nonumber\\
 &+\frac{1}{2} (\delta (x+t) +\delta (x-t) ) ]
 \label{term2}
 \end{align}
where the $\delta-$functions come from the term $\cos (qt) $. Note, this is actually the Fourier transform of $T_2$ 
 Now,
 \begin{align}
 \int_{0}^{t}\cosh (\tau\sqrt{1-q^2}) d\tau=\frac{\sinh (t\sqrt{1-q^2}) }{\sqrt{1-q^2}} .
 \end{align}
 Thus if we integrate Eq.~\eqref{term2} from $0$ to $t$, we get the Inverse Transform of the $T_3$ term. Using, $I'_0 (x) =I_1 (x) $ to do the integral, we get
 \begin{align}
  \mathcal{F}^{-1}[\frac{\sinh (t\sqrt{1-q^2}) }{\sqrt{1-q^2}}]=\frac{1}{2}I_{0} (\sqrt{t^2-x^2}) \Theta (t-|x|). 
  \label{term3}
 \end{align}
 Only the inverse of $T_1$ remains to be evaluated. We integrate l.h.s. of Eq.~\eqref{term3} from $0$ to $t$, to get
 \begin{align}
 \int_{-\infty}^{\infty}e^{-iqx}\frac{1}{1-q^2}\bigg (\cosh (t\sqrt{1-q^2}) -1\bigg) dq.
 \end{align}
 Taking derivatives with respect to $x$ twice, we get,
 \begin{align}
  -\int_{-\infty}^{\infty}e^{-iqx}\frac{q^2}{1-q^2}\bigg (\cosh (t\sqrt{1-q^2}) -1\bigg) dq
   \end{align}
   which is exactly twice the negative of $T_1$.  Thus, we need to do this same set of operations on the r.h.s. of Eq.~\eqref{term3} to get the inverse of the first term. Thus we have,
 \bea
 \cal{F}^{-1}[T_1]&=-\frac{1}{2}\frac{\partial^2}{\partial x^2}\bigg[\Theta (t-| x|) \int_{0}^{t}d\omega I_0 (\sqrt{\omega^2-x^2}) \Theta (\omega-| x |) \bigg]\nonumber\\
 &=\frac{1}{2}\frac{\partial}{\partial x}\bigg[\delta (t-| x|)  sgn (x) \int_{0}^{t}d\omega I_0 (\sqrt{\omega^2-x^2}) \Theta (\omega-| x |) \nonumber\\
& -\Theta (t-| x|) \frac{\partial}{\partial x}I_0 (\sqrt{\omega^2-x^2}) \Theta (\omega-| x |) \bigg].
\label{finverintermediate}
 \eea
 Now, because of the $\delta-$function, the first term on r.h.s. of the above equation is non-zero only when $|x|=t$, but then again putting that in the $\theta-$function, we get $0$, since $ (\omega-| t|) $ is always less than 0. So the first term on the r.h.s. is always zero. Thus the r.h.s. of Eq.~\eqref{finverintermediate} reduces to
 \begin{align}
 &-\frac{1}{2}\frac{\partial}{\partial x}\bigg[\Theta (t-| x|) \int_{0}^{t}\bigg (d\omega \frac{\partial I_0 (\sqrt{\omega^2-x^2}) }{\partial x}\Theta (\omega-| x|) \nonumber\\
&-I_0 (\sqrt{\omega^2-x^2}) \delta (\omega-| x|) sgn (x) \bigg) \bigg].
\label{finverinter2}
 \end{align}
 Doing the delta-function integral, \ie, the second integral above, we get $\Theta (t-| x|) sgn (x) $. The derivative of theta function in the first term gives zero in exactly the same way as above. Thus \eqref{finverinter2} becomes,
 \begin{align}
 &-\frac{1}{2}\Theta (t-| x|) \int_{0}^td\omega\frac{\partial^2 I_0 (\sqrt{\omega^2-x^2}) }{\partial x^2}\Theta (\omega-| x|) \nonumber\\ &+\frac{1}{2}\Theta (t-| x|) \int_{0}^td\omega\frac{\partial I_0 (\sqrt{\omega^2-x^2}) }{\partial x}\delta (\omega-| x|) sgn (x) \nonumber\\&-\frac{1}{2}\delta (t-| x|) +\frac{1}{2}\Theta (t-| x|) 2\delta (x) .
 \label{finverinter3}
 \end{align}
The second integral in \eqref{finverinter3} can be evaluated exactly and yields, $\frac{x}{\sqrt{\omega^2-x^2}}I_1 (\sqrt{\omega^2-x^2}) \delta (\omega-| x|) $.
 
  Now, using the properties of $I_1(z)$ for $z \to 0$, we write
 \begin{align}
 \lim _{\omega\rightarrow | x |}\frac{I_1 (\sqrt{\omega^2-x^2}) }{\sqrt{\omega^2-x^2}}=\frac{1}{2},\nonumber
 \end{align}
 and so the second term in \eqref{finverinter3} reduces to be $-\frac{|x|}{2}\Theta (t-| x|) $.\\
 Thus, 
 \begin{align}
  &\mathcal{F}^{-1}\bigg[\frac{q^2}{1-q^2}\bigg (\cosh (t\sqrt{1-q^2}) -1\bigg) \bigg]\nonumber\\&=-\frac{1}{4} (\delta (x+t) +\delta (x-t) ) +\frac{1}{2}\delta (x) -\frac{| x|}{8}\Theta (t-| x|) \nonumber\\&-\frac{1}{4}\Theta (t-| x|) \int_{0}^{t}d\omega \frac{\partial ^2}{\partial x^2}I_{0} (\sqrt{\omega^2-x^2}) \Theta (\omega-| x|) .
  \label{term1}
 \end{align}
 Thus, combining Eqs.~\eqref{term2}, \eqref{term3} and \eqref{term1}, we get the full inverse transform as written in the main text.

\section{No Flip Contribution for the Continuous Process}\label{nojump}

To calculate the position distribution for the continuous model in Sec.~\ref{sec:cont}, we have first subtracted the contribution from the trajectories with no $\sigma$-flips. In this section we calculate that contribution explicitly. If the particle starts at an angle $\theta_0$ (\ie, along $\hat{n}=\cos\theta_0\hat{x}+\sin\theta_0 \hat{y}$)  and does not undergo any change in the orientation till time $t$, then 
\bea
x=v_0t\cos\theta_0 \cr
y=v_0t\sin\theta_0. 
\eea
Thus contribution of this event to the probability distribution is
\bea
P_0(\vec r,t|\theta_0)= e^{-\gamma t}\delta (x-v_0\cos\theta_0 t) \delta (y-v_0\sin\theta_0 t). \label{eq:P0_theta0}
\eea 

To express it in terms of the radial coordinate $r$, we first take a Fourier transform of Eq.~\eqref{eq:P0_theta0} w.r.t. $\vec r \to \vec k$ and then integrate over the initial orientation $\theta_0.$  We  get,
\bea
\tilde P_0(\vec k,t) = e^{-\gamma t}J_0 (kv_0 t) 
\eea
where $k=\sqrt{k_1^2+k_2^2}$. Note that a Laplace transformation of the above expression w.r.t. $t \to s$ leads to $f(\vec k,s)$ given in Eq.~\eqref{EQ:fks} in the main text. 

To calculate $P_0(\vec r,t)$ we now take an inverse Fourier Transform from $\vec{k}\rightarrow\vec{r},$
\bea
P_0 (\vec r,t) &=&\frac{e^{-\gamma t}}{ (2\pi) ^2}\int_0^{\infty}k dk\int_0^{2\pi}d\psi e^{ikr\cos\psi}J_0 (kv_0 t) \nonumber\\
&=&\frac{e^{-\gamma t}}{2\pi}\int_0^{\infty}k dk J_0 (kr) J_0 (kv_0t). \n
\eea
which clearly depends only on the radial coordinate $r=\sqrt{x^2+y^2}.$ Now, we can use the identity from Section 6.512, Eq. 8 of \cite{Gradshteyn},
\bea
\int_{0}^{\infty}kJ_n (ka) J_n (kb) dk=\frac{1}{a}\delta (b-a) 
\label{EQ:BESSEL1}
\eea
and get,
\bea
P_0 (r,t) &=&\frac{e^{-\gamma t}}{2\pi r}\delta (r-v_0t) \n
\eea
which is quoted in Eq.~\eqref{eq:_P0rt} in the main text.	
	\section{Laplace Fourier Inversion of $\cal{G}(k,s)$ of the Continuous Process in the main text}\label{fourlapinver}
 We start from Eq.~\eqref{calgks} in the main text, and putting Eqs.~\eqref{EQ:fks} and \eqref{EQ:gfcontinuouslambdaks} in it, we have
 \bea
 \cal{G} (k,s) &=&\frac{\gamma}{ \left(\sqrt{ (s+\gamma) ^2+v_0^2 k^2}-\gamma\right) \sqrt{ (s+\gamma) ^2+v_0^2 k^2}}.~~
 \eea 
 \\
 Let us put $s+\gamma=s'$ and rewrite $\cal{G} (k,s) $ as
  \bea
 \cal{G} (k,s') &=&\frac{\gamma}{\left (\sqrt{s'^2+v_0^2 k^2}-\gamma\right) \bigg (\sqrt{s'^2+v_0^2 k^2}\bigg) }.
 \eea
 We can now take the $2-$d inverse Fourier transform from $\vec{k}$ to $\vec{r}$
 \bea
 \tilde{P} (r,s') &=&\frac{1}{ (2\pi) ^2}\int_{0}^{\infty}k dk \int_{0}^{2\pi}d\psi e^{ikr\cos (\psi) }\cal{G} (k,s') \n
 \eea
 where $\psi$ is the angle between $\vec{k}$ and $\vec{r}$.  Doing the $\psi$ integral,
 \bea
 \tilde{P} (r,s') &=&\frac{1}{2\pi}\int_{0}^{\infty}k~dk~J_0 (kr) \cal{G} (k,s'). 
\label{EQ:cont_rlaplace}
 \eea
 Doing the $k$ integral is non-trivial. We first use an integral identity \cite{Gradshteyn}
 
 \bea
 \int_{0}^1 d w\frac{ (a-w) }{[ (a-w) ^2+b^2 (1-w^2) ]^{3/2}}&=&\frac{1}{\sqrt{a^2+b^2} (\sqrt{a^2+b^2}-1) }. \cr
 && \label{intidentity1} 
 \eea

 The right-hand side of the above identity can be mapped to $\cal{G} (k,s')$ by identifying  $a=\frac{s'}{\gamma}$ and $b=\frac{kv_0}{\gamma}$. Thus, we can write,
 \bea
\cal{G} (k,s')= \gamma\int_0^1 dw \frac{ (s'-\gamma w) }{[ (s'-\gamma w) ^2+k^2v_0^2 (1-w^2) ]^{\frac{3}{2}}}.
\label{Gks_after1}
 \eea
 We now use an integral Bessel Function identity from Section 6.611 Eq. 1 of \cite{Gradshteyn},
 \bea
 \int_0^{\infty}dt e^{-\alpha t}tJ_0 (\beta t) =\frac{\alpha}{ (\alpha^2+\beta^2) ^\frac{3}{2}}.\nonumber
 \eea
 Again the right-hand side of this identity can be mapped to the integrand in Eq.~\eqref{Gks_after1} if  $\alpha= (s'-\gamma w) $ and $\beta=kv_0\sqrt{1-w^2}$. Thus $\cal{G} (k,s') $ becomes
 \bea
 \gamma\int_{0}^1 dw\int_{0}^{\infty}dt e^{- (s'-\gamma w) t}t J_0 (kv_0t\sqrt{1-w^2}). 
 \eea
 Putting this back in the expression for $\tilde{P}(r,s'),$ \ie, Eq.~\eqref{EQ:cont_rlaplace}, and substituting back $s=s'-\gamma,$ we have
  \bea
 \tilde{P}(r,s) &=& \frac{\gamma e^{-\gamma t}}{2\pi} \int_{0}^{\infty}k~ d k \int_{0}^{\infty}dt~e^{-st}t \int_{0}^{1}d w ~e^{\gamma w t} \cr
  &\times & J_0 (kr) J_0 (k v_0 t\sqrt{1-w^2}). 
 \eea
 Thereafter doing the k integral, we have
 \bea
\tilde{P}(r,s) &=& \frac{\gamma e^{-\gamma t}}{2\pi v_0}\int_{0}^{\infty}dt~e^{-st} \int_{0}^{1}dw~ e^{\gamma w t} \frac{\delta (r-v_0 t\sqrt{1-w^2}) }{\sqrt{1-w^2}}~~~~~~~
 \eea
Since the above equation is already in the form of a Laplace transformation $\int_0^{\infty} d t~ P(r,t) e^{-s t}$, the inverse transform $P(r,t)$ can be immediately read out,
 \bea
 P (r,t) &=&\frac{\gamma e^{-\gamma t}}{2\pi v_0}\int_{0}^{1}dw \frac{\delta (r-v_0 t\sqrt{1-w^2}) }{\sqrt{1-w^2}}e^{\gamma w t}.\nonumber
\eea 
 The $w$-integral can be done immediately due to the presence of the $\delta$-function and yields, 
\bea
 P (r,t) &=&\frac{\gamma e^{-\gamma t}}{2\pi v_0}\frac{\exp\bigg[\frac{\gamma}{v_0} \sqrt{v_0^2 t^2-r^2}\bigg]}{\sqrt{v_0^2 t^2-r^2}}.
 \eea

\section{Large Deviation Function for the Continuous Case from the Generating Function using Saddle Point Approximation}

Here we show how to get the large deviation form for the marginal distribution in the continuous case, without inverting the generating function exactly.\\
The generating function in $s$ space, $\int_0^{\infty} dt~ e^{-st}\la e^{k x}\ra$, has the same form in $1-$d as Eq.~\eqref{EQ:gfcontinuouslambdaks} in the main text, with $\vec{k}$ being replaced by $1-$d vector $k$. Let us denote it by $G(k,s)$
\bea
G (k,s) &=&\frac{1}{\sqrt{ (\gamma +s) ^2+k^2 v_0^2}-\gamma}.
\eea
We want to invert $G (k,s) $ with respect to $s$ first, to obtain the generating function. To do that we write the Bromwich integral as follows
 \bea
  \tilde{P} (k,t) &=&\frac{1}{2\pi i}\int_{-i\infty +c}^{i\infty +c}\frac{ds~ e^{s t}}{\sqrt{ (\gamma +s) ^2+k^2 v_0^2}-\gamma}.\nonumber
\eea
It is straightforward to see that the integrand has two simple poles at $-\gamma \pm\sqrt{\gamma^2-k^2 v_0^2}$ and two branch points at $-\gamma \pm ik v_0$. 
At large times, the integral is dominated by the contribution  from the pole closet to the origin $-\gamma +\sqrt{\gamma^2-k^2 v_0^2},$ and we can write,
 \bea
 \tilde{P} (k,t) &\approx &\frac{\gamma e^{-\gamma t +t\sqrt{\gamma ^2-k^2 v_0^2}}}{\sqrt{\gamma^2-k^2 v_0^2}}.
 \label{gfcontinuous}
 \eea
In this large time limit, the position distribution is then given by,

\bea
  P (z=\frac{x}{v_0 t},t) &=&\gamma e^{-\gamma t}\int_{-\infty}^{\infty}dk\frac{e^{v_0 t (-i k z+\sqrt{k^{2}  +\frac{\gamma^2}{v_0^2}}) }}{\sqrt{k^{2} v_0^2+\gamma^2}}.\nonumber
\eea
  This integral can be computed  using the saddle point approximation. Let us denote $g (k,t) =v_0 t (-k z+\sqrt{k^{2}  +\frac{\gamma^2}{v_0^2}})$ which has a maximum at $k_*$ satisfying $g' (k_*) =0.$ Keeping terms up to second order about the maximum, the saddle point integral comes out to be 
  \bea
 P (z,t) &=& \gamma e^{-\gamma t}\frac{e^{g (k_*) }}{\sqrt{k_*^{2} v_0^2+\gamma^2}}\frac{1}{\sqrt{2\pi g'' (k_*) }}\nonumber\\
 &=& \frac{e^{-\gamma t (1-\sqrt{1-z^2}) }}{\sqrt{4\pi D_\text{eff}t (1-z^2) ^{\frac{1}{2}} }}.
 \eea

\end{document}